\documentclass[10pt,final,noinfoline]{imsart}

\usepackage[a4paper, total={6in, 9in}]{geometry}

\usepackage{amssymb}
\usepackage{amsmath}
\usepackage{amsthm}

\usepackage{url} 

\theoremstyle{plain}
\newtheorem{theorem}{Theorem}[section]

\theoremstyle{definition}
\newtheorem{definition}{Definition}[section]

\theoremstyle{remark}

\newtheorem{remark}{Remark}

\usepackage{graphicx}
\graphicspath{ {./graphics/} }
\usepackage{algorithm}
\usepackage{algpseudocode}
\usepackage{tikz}
\foreach \a in {q,w,e,r,t,y,u,i,o,p,a,s,d,f,g,h,j,k,l,z,x,c,v,b,n,m,Q,W,E,R,T,Y,U,I,O,P,A,S,D,F,G,H,J,K,L,Z,X,C,V,B,N,M}{
  \expandafter\xdef\csname v\a\endcsname {
		{\noexpand\mathbf{\a}}
	}
}

\newcommand{\vone}{{\mathbf{1}}}
\newcommand{\vzero}{{\mathbf{0}}}
\newcommand{\vtheta}{{\boldsymbol{\theta}}}

\newcommand{\Aa}[1]{($\mathcal{A}#1$)}
\newcommand{\Asa}{\Aa{1}}
\newcommand{\Asb}{\Aa{2}}
\newcommand{\Asc}{\Aa{3}}

\usepackage{makecell}

\usepackage{natbib}
 \bibpunct[, ]{(}{)}{,}{a}{}{,}%
 \def\bibsep{\smallskipamount}%
\begin{document}

\begin{frontmatter}



\title{Modeling Innovation Ecosystem Dynamics through Interacting Reinforced Bernoulli Processes} %% Article title
\runtitle{Modeling Innovation Dynamics}


\author{\fnms{Giacomo} \snm{Aletti}\ead[label=e1]{giacomo.aletti@unimi.it}},
\author{\fnms{Irene} \snm{Crimaldi}\ead[label=e2]{irene.crimaldi@imtlucca.it}},
\author{\fnms{Andrea} \snm{Ghiglietti}\ead[label=e3]{andrea.ghiglietti@unimib.it}}
\and
\author{\fnms{Federico} \snm{Nutarelli}%\thanksref{t1}
\ead[label=e4]{federico.nutarelli@imtlucca.it}}

% \thankstext{t1}{Corresponding author.}
\runauthor{G. Aletti, I. Crimaldi, A. Ghiglietti, F. Nutarelli}

\address{G. Aletti\\
Department of Environmental Science and Policy\\
University of Milan\\
Via Celoria 2, 20133\\
Milano, Italy\\
\phantom{E-mail:\ giacomo.aletti@unimi.it}}

\address{I. Crimaldi, F. Nutarelli\\
IMT School for Advanced Studies Lucca,\\
Piazza San Ponziano 6, 55100\\
Lucca, Italy\\
\phantom{E-mail:\ irene.crimaldi@imtlucca.it, federico.nutarelli@imtlucca.it}}

\address{A. Ghiglietti\\
Department of Statistics and Quantitative Methods\\
Università degli Studi di Milano-Bicocca\\
Via Bicocca degli Arcimboldi 8\\
Milan, Italy\\
\phantom{E-mail:\ andrea.ghiglietti@unimib.it}}

%% Abstract
%% Abstract
\begin{abstract}
Innovation is cumulative and interdependent: successful inventions build on prior knowledge within technological fields and may also affect success across related ones. Yet these dimensions are often studied separately in the innovation literature. This paper asks whether patent success across technological categories can be represented within a single dynamic framework that jointly captures within-category reinforcement, cross-category spillovers, and a set of aggregate regularities observed in patent data. To address this question, we propose a model of interacting reinforced Bernoulli processes in which the probability of success in a given category depends on past successes both within that category and across other categories. The framework yields joint predictions for success probabilities, cumulative successes, relative success shares, and cross-category dependence.\\
\indent We implement the model using granted US patent families from GLOBAL PATSTAT (1980--2018), defining category-specific success through a cohort-normalized forward-citation index. The empirical analysis shows that successful innovations continue to accumulate, but less than proportionally to the growth in patent opportunities, while technological categories remain interdependent without becoming homogeneous. Under a mean-field restriction, the model-based inferential exercise yields an estimated interaction intensity of $0.643$, pointing to positive but non-maximal interaction across technological categories.
\end{abstract}

%% Keywords
\begin{keyword}
Innovation Ecosystems \sep Technological Interdependence \sep Interacting Bernoulli Processes \sep Patent Interactions \sep Technological Specialization
\end{keyword}

\end{frontmatter}

\section{Introduction}\label{sec:intro}

Innovation is cumulative and interdependent. Innovative success builds on prior knowledge within technological fields, but it is also shaped by spillovers across related fields. These two ideas run through much of the innovation literature. Early work emphasized the role of research, technological opportunity, and appropriability conditions in shaping inventive performance \citep{griliches1958research, evenson1968contribution, levin1985appropriability, jaffe1986technological}. Subsequent contributions showed that innovative activity also differs systematically across sectors because of differences in cumulativeness, demand conditions, and sectoral regimes \citep{malerba1996schumpeterian, malerba2002sectoral}. A related literature documented intersectoral technological linkages and knowledge spillovers, highlighting how inventive activity in one domain may affect innovation in other technologically related domains \citep{rosenberg1979technological, scherer1982inter, jaffe1989characterizing}. More recently, work on recombinant innovation and on the ``adjacent possible'' has stressed that innovation often proceeds through combinations of related knowledge elements and expansion into nearby technological opportunities \citep{fleming2001recombinant, kauffman2000investigations, monechi2017waves, taalbi2020evolution}. At the same time, a long tradition --- from early concerns about the rate of invention to recent evidence on declining disruptiveness and research productivity --- suggests that producing influential innovations may become more difficult as the stock of existing knowledge expands \citep{stafford1952rate, schmookler1954level, jones1995rd, jones1995time, ha2007accounting, park2023papers, madsen2024declining}.\\

What is less developed in the innovation literature is a parsimonious dynamic framework that studies these elements within a single empirical setting. The literature has documented cumulativeness within fields, spillovers across fields, and broad regularities in innovative outcomes, but these dimensions are often examined separately or through reduced-form correlations, proximity measures, and network mappings rather than through a common dynamic model of innovation success \citep{jaffe1989characterizing, pichler2020technological, taalbi2020evolution, colladon2025new}. This paper addresses that issue by asking whether innovation success across technological categories can be represented within a single stochastic framework that jointly captures within-category reinforcement, cross-category spillovers, and a small set of aggregate patterns observed in patent data.

We propose such a framework. Specifically, we model category-specific patent successes as a system of interacting reinforced Bernoulli processes. At each patent time-step, a category-specific indicator takes value one if the patent qualifies as a success in that category and zero otherwise. The probability of success in category $h$ depends on past successes in the same category and on cumulative successes in other categories. Cross-category dependence is summarized by a non-negative interaction matrix $\Gamma=(\gamma_{j,h})$, which should be interpreted as a reduced-form representation of technological linkages rather than as a model of how such linkages themselves emerge. In this sense, the paper does not endogenize the formation of technological relations; rather, it studies the dynamic implications of reinforcement and interdependence for the evolution of innovation success across categories.

The empirical application uses granted US patent families from GLOBAL PATSTAT, classified at the CPC-1 level. Our empirical object is not patenting per se, nor novelty in the legal sense used in patent examination. Rather, it is success in a technological category, measured ex post through technological influence on subsequent inventions. This distinction matters. In the innovation literature, novelty and radicalness are broader concepts than legal novelty and are often associated with recombination, distinctiveness, and impact \citep{fleming2001recombinant, dahlin2005when}. In the present paper, however, we focus on success as the observable outcome to be modeled dynamically. As baseline outcome, we define a category-specific patent success through a cohort-normalized forward-citation measure following \citet{squicciarini2013measuring}: for a given patent and target category, we count subsequent citations from patents in that category within a fixed time window and normalize by the maximum value observed in the patent's year-by-category cohort. This normalization is intended to improve comparability across cohorts and technological classes. Moreover, in order to neutralize any truncation effect in citations, we pay attention to the fact that all the patents considered for the empirical analysis share the same time-window to receive citations, but it does not mechanically impose the dynamic patterns studied later in the paper. Indeed, it is a within-cohort rescaling, whereas the theoretical results concern the sequential accumulation of successes across patents in system time. In other terms, the normalization makes patents more comparable across cohorts, but the dynamical properties of the model arise from the reinforcement mechanism rather than from the normalization itself.

This empirical construction yields, for each patent and category, a binary success indicator that can be interpreted as the observable counterpart of the success events studied in the model. The relevant time dimension is \emph{system time}: the model concerns the conditional probability that the next observed patent is a success in category $h$, given the history of previously observed patents. This point is important for interpretation. When the model implies declining success probabilities, it does not refer to a purely mechanical trend in calendar time. Rather, it means that, as the number of patent opportunities already observed increases, cumulative successes grow less than proportionally. In this sense, success becomes rarer \emph{per patent opportunity}, not because a downward trend is imposed ex ante, but because it is implied by the reinforcement dynamics. The latter pattern is consistent with the idea that influential innovations may become harder to generate as the system expands and the knowledge burden increases \citep{bloom2020ideas, jones2009burden}. \\~\\
\indent The paper contributes to innovation research by proposing a dynamic stochastic framework for category-specific innovation success that jointly captures cumulativeness within technological categories and spillovers across categories. The framework yields three aggregate implications that can be taken to the data: cumulative successes grow sublinearly in system time; relative success shares across categories stabilize in the long run; and cross-category comovement weakens over time in the empirically relevant unbalanced case. We compare these implications with patent data and complement this comparison with a parsimonious inference exercise on interaction intensity under a mean-field restriction. Section \ref{sec:rob_checks} in the Appendix performs an econometric analysis as a robustness check to confirm  whether success in a target category is positively associated with past success in the same category (path dependence) and with past success in the other categories (cross-fertilisation).\\~\\
\indent The remainder of the paper is organized as follows. Section~\ref{sec:cfc} defines the category-level success measure and discusses its interpretation. Section~\ref{sec:mathbkg} presents the interacting Bernoulli framework and its main theoretical implications. Section~\ref{sec:data_analysis} examines whether the aggregate patterns predicted by the model are present in the patent data and reports the complementary inference exercise on interaction intensity. Section~\ref{concl} concludes.

\section{Category forward citation index and definition of ``success''} \label{sec:cfc}

In this paper, patents are used as observable innovation outcomes. Our empirical object is not patenting per se, but whether a patent qualifies as a \emph{successful} innovation in a given technological category according to an ex post citation-based criterion. This distinction is important. In patent examination, novelty is a legal requirement: a patent must be new relative to the prior art at the time of filing. In the innovation literature, however, novelty and radicalness are broader concepts, often linked to recombination, distinctiveness, and subsequent impact \citep{fleming2001recombinant, dahlin2005when}. redOur empirical measure does not aim to operationalize novelty in all these senses. Rather, it focuses on \emph{success} as an observable ex post measure of technological influence. A patent may therefore be legally novel and yet have limited technological impact, while a highly cited patent may indicate an invention that proved especially important for later developments. For this reason, throughout the paper we reserve the term ``success'' for patents that exceed a category-specific threshold based on forward citations.

Forward citations are widely used in the innovation literature as indicators of the technological importance of patents for subsequent developments \citep{dahlin2005when, squicciarini2013measuring}. We therefore build on the forward-citation index proposed by \citet{squicciarini2013measuring}, adapting it to a category-specific setting. The empirical construction is not tied to a specific classification system or to this particular index: the same framework could be implemented with alternative taxonomies or alternative measures of technological influence.\footnote{A review of novelty and impact measures that could be used in our setting is provided in Appendix~\ref{appx:litindx}.}

We consider a finite set $\mathcal H$ of technological categories. For each patent $n$, we denote by $c_n$ its source category in the empirical classification and by $d_n$ its publication (full) date. Success, however, is evaluated separately for each target category $h\in\mathcal H$. Hence, a patent is associated with one source category $c_n$ but may qualify as a success in none, one, or several target categories, depending on the citations it subsequently receives from patents in those categories. This distinction between source category and target-category success is central to the empirical construction used in the rest of the paper.

At the empirical level, technological categories are defined at the CPC-1 level. On the citing side, category-specific citation counts may reflect all technological categories represented in the citing patent, so the target-category dimension can incorporate multiple CPC-1 classes.\footnote{Approximately 35\% of the patents in the dataset belong to multiple CPC-1 classes.} By contrast, whenever a unique source category is required in the empirical analysis below, each patent is assigned to a single CPC-1 source category, defined as the modal first-level CPC letter among the CPC codes associated with that patent in the original data.\footnote{This source-category label is used in the construction of the origin-based quantities discussed later in the paper. The corresponding analysis therefore relies on a unique source-category assignment rather than on fractional allocation across multiple source categories. This should be read as a pragmatic data-reduction rule adopted for empirical tractability.}

For a given patent $n$, with publication date $d_n$, and for each category $h\in\mathcal H$, let
\[
CIT_{n,h}
\]
denote the number of patents published in the period $[d_n,d_n+365\times T]$, belonging to category $h$, and citing patent $n$.

Now consider patent $n$, published in year $y_n$ and belonging to source category $c_n$. We refer to the set of patents published in the same year $y_n$ and belonging to the same source category $c_n$ as the \emph{cohort of patent $n$}. For each target category $h$, we define the index
\[
I_{n,h}=\frac{CIT_{n,h}}{\max_{\{i \in \text{cohort of } n\}} CIT_{i,h}}\,.
\]
Thus, $I_{n,h}$ measures the forward citations received by patent $n$ from category $h$, normalized by the maximum value observed within the patent's year-by-source-category cohort.

Note that, in practical application, if the available dataset contains patents published until a certain year $T_f$, after we have chosen the number of years $T$ to be considered for the computation of the index, we perform our analysis for each patent belonging to the available dataset with publication year before or equal to $(T_f-T)$ so that,  
 for each considered patent, we  have the entire time-window $[d_n,d_n+365\times T]$ required for the computation of the index. This is sufficient to neutralize any truncation effect in citations, since all the considered patents have the same time-window to receive citations.  Moreover, the purpose of the normalization in $I_{n,h}$ is the following. It accounts for the fact that forward citations vary substantially across publication years and technological categories, making raw citation counts difficult to compare across cohorts. The normalization is computed within the patent's year-by-source-category cohort, in order to  mitigate comparability problems in citation counts. However, the normalization does not mechanically impose the dynamic patterns studied later in the paper: it is a within-cohort rescaling, whereas the theoretical results concern the sequential accumulation of successes across patents in system time. In other words, the normalization makes patents more comparable across cohorts, but the asymptotic properties of the model arise from the reinforcement dynamics rather than from the normalization itself.

This category-specific index is a disaggregation of the forward-citation index in \citet{squicciarini2013measuring}. Indeed, if $I_S(n)$ denotes their index, then
\[
I_S(n)=\frac{CIT_n}{\max_{\{i \in \text{cohort of } n\}} CIT_i}
=
\sum_{h\in \mathcal H}
\frac{\max_{\{i \in \text{cohort of } n\}} CIT_{i,h}}
{\max_{\{i \in \text{cohort of } n\}} CIT_i}
\, I_{n,h}.
\]

\begin{remark}[Invariance property of the index]
Note that, if we restrict our attention to (or we can observe) only the patents published in a certain category $h^*$, then, for each patent $n$ belonging to this category, the index value  $I_{n, h^*}$ remains well defined and its value does not change with respect to the one computed if we consider all the published patents. Indeed, its definition is based only on the observation of the patents belonging to category $h^*$.
 \end{remark}

By construction, $I_{n,h}\in[0,1]$. To obtain the binary success indicators required by the model, we fix a threshold $\tau$ and classify a patent as a success in category $h$ whenever its index exceeds that threshold \footnote{The threshold does not determine which patents enter the sample. Rather, it determines whether each patent-category observation is coded as a success or a failure.}. The threshold therefore translates a continuous measure of technological influence into a success/failure outcome. Different threshold choices correspond to different degrees of selectivity in the definition of success; for this reason, the choice of $\tau$ and its empirical implications are discussed in the supplementary material and revisited in the data analysis. 

\begin{definition}[Success in category $h$]
Given a threshold $\tau$, patent $n$ is said to be a success in category $h$ if
\[
I_{n,h}>\tau.
\]
\end{definition}

Accordingly, let $X_{n,h}$ be the binary variable equal to $1$ if patent $n$ is a success in category $h$ and equal to $0$ otherwise. This yields a finite system of Bernoulli processes,
\[
\{X^{(h)}=(X_{n,h})_n : h\in\mathcal H\},
\]
which provides the empirical counterpart of the success events modeled in the next section. These indicators are not assumed to be independent. Rather, they are the observable category-specific success events whose dynamics we seek to represent through within-category reinforcement and cross-category interdependence.

In the next section, we introduce the dynamic stochastic framework and derive the aggregate implications that will later be compared with the patent data. Although the empirical implementation developed here is based on CPC categories and a citation-based success measure, the same methodological approach can be adapted to alternative classifications and alternative success indicators.

%\section{Mathematical model and theoretical results: Interacting Bernoulli processes} \label{sec:mathbkg}

\section{A dynamic framework for innovation success across categories}\label{sec:mathbkg}

This section develops the evolution stochastic model which is at the core of the paper. Starting from category-specific success indicators, such as the ones introduced in Section \ref{sec:cfc}, we describe how innovation success evolves when it is both cumulative within technological categories and interdependent across related categories. The purpose of the model is to characterize the dynamic implications of reinforcement and cross-category interdependence for the evolution of category-specific innovation success over time.\\
In the patent application considered in this paper, each process refers to a technological category and the time-step $t$ indexes patents sequentially according to publication date. Unless otherwise specified, however, the model is presented in a general form and may be applied in other contexts as well.

\indent Let $\mathcal H$ be a finite set with $N=\text{card}(\mathcal{H})$. We are going to define a system $\{ X^{(h)}=(X_{t,h})_{t\in \mathbb{N}\setminus \{0\}} \,:\, h\in \mathcal{H} \}$ of  
Bernoulli processes with parameters evolving along time according to a reinforcement rule with interaction: specifically, for each $t\geq 0$, 
let $X_{t+1,1},\dots, X_{t+1,N}$ be $N$ Bernoulli random variables, that are conditionally independent given the past and such that  
\begin{equation}\label{eq-model}
\begin{aligned}
P_{t,h}&=P(X_{t+1,h}=1\,|\,\text{past until time-step $t$})
\\
&=
\frac{\theta_h + \sum_{j\in {\mathcal H}} \gamma_{j,h} S_{t,j}}{c+t}
\end{aligned}
\end{equation}
where $\theta_h>0$, $c\geq\theta_h$ are parameters that tune the initial condition in the model dynamics (for details, see \eqref{eq-dynamics-vector} in Appendix~\ref{app-proofs}) and $S_{t,j}=\sum_{n=1}^t X_{n,j}$ is the number of successes until time-step $t$ for process $j$. 

The parameters $\gamma_{j,h}$ play the fundamental role in the dynamics of the system (see again \eqref{eq-dynamics-vector}) and summarize how past successes in category $j$ affect future success in category $h$. We refer to the matrix $\Gamma=(\gamma_{j,h})_{j,h}$ as the interaction matrix. In the empirical interpretation of the model, $\Gamma$ should be read as a reduced-form representation of technological interdependence across categories. We further assume the following conditions:

\begin{itemize}
\item[{\Asa}] $\Gamma$ is non-negative and such that 
$\mathbf{1}^\top\Gamma\leq\mathbf{1}^\top$ (where $\vone = (1,\ldots,1)^\top $), i.e. 
$\sum_{j\in {\mathcal H}} \gamma_{j,h}\leq 1$ for each $h\in {\mathcal H}$ 
so that we have $P_{t,h}\in (0,1)$;
\item[{\Asb}] $\Gamma$ is irreducible, that is the graph with the Bernoulli processes as nodes and with $\Gamma$ as the (weighted) adjacency matrix is strongly connected. (In Appendix \ref{app-reducible-case} we will discuss the case of a reducible matrix $\Gamma$.\footnote{This extension may be particularly relevant in contexts where the USPTO --or other patent offices—modify the structure of technological categories through reclassification \citep{lafond2019long, chae2019study} thus changing the landscape of interaction among the latter.})
\end{itemize}

Assumption {\Asa} ensures that reinforcement does not make success probabilities explode beyond the unit interval, while Assumption {\Asb} guarantees that no category is isolated from the rest of the system. Together, these conditions describe a setting in which successes can propagate across categories -- consistently with the innovation literature emphasizing that technological outcomes are shaped by interdependence and spillovers across related domains rather than evolving in isolation \citep{jaffe1989characterizing, taalbi2020evolution}--  but in a statistically novel and well-defined way \footnote{Namely, our framework relates to the literature on reinforced stochastic processes surveyed by \citet{pemantle-2007}, which includes generalized P\'olya urns, interacting urn models, and other processes with reinforcement. Relative to that literature, our contribution is to formulate a multi-category system of patent-success processes with an economically interpretable interaction matrix and to derive joint predictions for success probabilities, cumulative successes, relative shares, cross-category dependence, and shock propagation.}. \\

The economic interpretation of the model is straightforward. A higher number of past successes in category $h$ raises the probability of future success in the same category. At the same time, past successes in other categories may also raise the probability of success in $h$, capturing cross-category spillovers and technological interdependence. The parameters $\gamma_{j,h}$ govern the strength of these direct cross-category effects.  In other words, given the sequence of patents observed up to time $t$, the probability that the next patent is a success in category $h$ depends on the cumulative record of past successes both within category $h$ and across other categories. In this sense, our framework provides a dynamic stochastic representation of how within-category cumulativeness and cross-category spillovers jointly shape the evolution of innovation success, capturing cumulativeness and path dependence within technological fields.

More formally, at each time-step $t$, the probability $P_{t,h}$ that we will have a success for process $h$ at time-step $t + 1$ can have an increasing dependence, not only on the number of successes observed in process $h$ itself, according to a {\em self-reinforcement principle} \citep{pemantle-2007}, but also on the number $S_{t,j}$ of successes observed in any other process $j$ until time-step $t$ (a property that we call {\em cross-reinforcement}). The parameter $\gamma_{j,h}$ regulates this dependence (with $\gamma_{j,h}=0$ meaning the absence of direct dependence). In other words, for each pair $(j, h)$ of processes, the parameter $\gamma_{j,h}$ quantifies how much the appearance of a success in process $j$ induces a potential future success in process $h$. It is important to note that the assumption of irreducibility of $\Gamma$ entails an {\em interaction} in this sense among all the processes: process $j$ can reinforce process $h$ directly (the case when $\gamma_{j,h}>0$) or indirectly, through the presence of at least one path joining them.\\

We first characterize the long-run behavior of success probabilities. This result is useful for interpretation, because it shows when influential innovations become rarer relative to the number of patent opportunities generated by the system.

\begin{theorem}\label{th-P} Under assumptions {\Asa} and {\Asb}, denote by $\gamma^* \in (0, 1]$ the
Perron-Frobenius eigenvalue of $\Gamma$. Then, for each $h\in{\mathcal H}$, we have as the number $t$ of patents grows
$$
t^{1-\gamma^*} P_{t,h}\stackrel{a.s.}\longrightarrow P_{\infty,h}\,,
$$
where $P_{\infty,h}$ is a finite strictly positive random variable. Moreover, for each $h,j\in{\mathcal H}$, the ratio $P_{\infty,h}/P_{\infty,j}$ of the above limit random variables is almost surely equal to the deterministic quantity $u_h/u_j>0$,  
where $\vu = (u_h)_h$ is the (unique up to a multiplicative non-zero constant) left eigenvector of $\Gamma$ associated to $\gamma^*$. Finally, the above convergence is also in quadratic mean. 
\end{theorem}

 We recall that -- according to the Perron-Frobenius theory -- the components of the vector $\vu$ are all different from zero and with the same sign and, in graph theory, they correspond to the relative eigenvector centrality scores (with respect to $\Gamma^\top$). 
 %This means, intuitively that a high value of $u_h$ means that success in category $h$ is more strongly shaped by past successes in other categories that occupy central positions in the system of technological interdependence. 
 \footnote{The eigenvector centrality is a measure of the importance of a node in a graph with respect to its out-links (if it is computed for the adjacency matrix) or its in-links (if it is computed for the transpose of the adjacency matrix). A high eigenvector centrality score means that the node points to, or respectively is pointed to by, many nodes with high scores. In the above theorems, $\vu$ is the vector of the relative eigenvector centrality scores with respect to $\Gamma^\top$ and, hence, by \eqref{eq-model}, a high value of $u_h$ means that the probability of having a success for process $h$ depends on the number of past successes observed in many processes that themselves have high scores.}.
Hence, the quantity $u_h$ is a (relative) measure of  how strongly category $h$ is exposed to successes coming from categories that are themselves central in the system of technological interdependencies, consistently with the idea that technological position and network linkages shape both spillovers and subsequent innovative outcomes \citep{jaffe1989characterizing, taalbi2020evolution}.\\
This result implies that
\begin{itemize}
\item in the case $\gamma^*<1$ (that is when $\vone^\top \Gamma\neq \vone^\top$) the probability of observing a success converges almost surely toward zero, i.e. successful innovations continue to occur, but they become progressively rarer relative to the number of patent opportunities observed in the system and this convergence happens sublinearly at the same rate $1/t^{(1-\gamma^*)}$ for every process (category) $h$; 
\item in the case $\gamma^*=1$ (that is when $\vone^\top \Gamma=\vone^\top$) the probability of observing a success in process $h$ converges almost surely to a finite strictly positive random limit, which is the same for each process (note that in this case $\vu$ is, up to a multiplicative non-zero constant, equal to the vector $\vone$); in other words, at the steady state, the probability of having a success in a process is the same for each process of the system. Economically, this means that cross-category reinforcement is strong enough to prevent the long-run success probability from vanishing.
\end{itemize}

\begin{remark} A clarification on the time scale is useful here. The quantity $P_{t,h}$ is defined in system time: it is the probability that the next patent in the sequence is a success in category $h$, conditional on the history of previously observed patents. Therefore, the statement that $P_{t,h}$ declines when $\gamma^*<1$ should not be read as a purely calendar-time trend. Rather, it means that, conditional on the number of patent opportunities already generated by the system, additional successes become progressively rarer. This is precisely why the result is non-trivial. The model does not start from an assumed decline in $P_{t,h}$. Instead, the decline is derived from the reinforcement dynamics and the interaction structure summarized by $\Gamma$. The asymptotic relation for $S_{t,h}$ implies that success probabilities decrease because cumulative successes grow less than proportionally with the number of observed patent opportunities.
\end{remark}

\indent This distinction is useful for interpreting the model in innovation terms. When $\gamma^*<1$, the system displays a form of subcritical reinforcement: past successes still matter, but not enough to sustain a constant rate of influential innovations as the system expands. When $\gamma^*=1$, instead, interdependencies across categories are sufficiently strong to keep success probabilities persistently positive in the long run. This interpretation is consistent with the literature showing, on the one hand, that the absence of autocatalytic structures can weaken innovation dynamics \citep{napolitano2018technology} and, on the other hand, that cross-domain knowledge flows and technological linkages can sustain innovation opportunities across domains \citep{jaffe1989characterizing, kim2017dynamic, taalbi2020evolution}. As we will see in what follows, our data analysis points to the empirically relevant case $\gamma^*<1$.\\

The most relevant implications for the empirical analysis concern the observable cumulative counts of successes. Unlike the conditional probabilities $P_{t,h}$, these quantities can be directly constructed from the patent data and therefore provide the main bridge between the model and the empirical patterns studied in Section~\ref{sec:data_analysis}.

\indent In particular, from Theorem~\ref{th-P}, we can also deduce the following result for the processes 
$(S_{t,h})_t$, with $h\in{\mathcal H}$, that represent the number of successes observed along time in 
each Bernoulli process $h$. We enforce here again that, differently from the previous quantities $P_{t,h}$, that are not observable (due to the presence of the unknown model parameters), the quantities $S_{t,h}$ {\em can be directly observed} and so used for data analyses. 

\begin{theorem}\label{th-S} Under assumptions {\Asa} and {\Asb}, denote by $\gamma^* \in (0, 1]$ the
Perron-Frobenius eigenvalue of $\Gamma$ and by $\vu = (u_h)_h$ its corresponding (unique up to a multiplicative non-zero constant) left eigenvector. Then, for each $h\in{\mathcal H}$, we have 
$$
\frac{S_{t,h}}{t^{\gamma^*}}\stackrel{a.s.}\longrightarrow S_{\infty, h}  \,,
$$
where $S_{\infty, h}$ is a finite strictly positive random variable. Moreover, 
for each $h,j\in {\mathcal H}$, we have 
$$
\frac{S_{t,h}}{S_{t,j}}\stackrel{a.s.}\longrightarrow \frac{u_h}{u_j}
\quad\text{and}\quad
\frac{S_{t,h}}{\sum_{j=1}^N S_{t,j}}\stackrel{a.s.}\longrightarrow \frac{u_h}{\sum_{j=1}^N u_j}\,.
$$
\end{theorem}

This result states that the number $S_{t,h}$ of successes for all the processes $h$ grows with the same Heaps’ exponent $\gamma^*\in (0,1]$: that is $S_{t,h}\stackrel{a.s.}\sim S_{\infty,h}\, t^{\gamma^*}$. In addition, the ratio
$S_{t,h} / S_{t,j}$ provides a strongly consistent estimator of the ratio $u_h /u_j$ of the relative eigenvector centrality scores (with respect to $\Gamma^\top$) of the two nodes (processes) $h$ and $j$ and the share of successes observed for process $h$ converges almost surely to the absolute eigenvector centrality score of $h$. \\

\indent It is interesting to note that, in the innovation framework, Theorem~\ref{th-S} highlights how the long-run proportions of successful patents in each category reflect the connectedness and influence of that category within the broader innovation network. While this relationship is formally established in the present work, related insights have been suggested in the literature without formal proof. For example, \citet{pichler2020technological} show that the innovation rate of a technological domain is influenced by the innovation rates of the domains it depends on. Similarly, \citet{sampat2002cite, zhang2025whom} emphasize the importance of the quality of a patent's prior connections as well as their technological domain.\\

Because the framework is stochastic, it also yields implications for fluctuations and cross-category comovement. We believe that the latter additional implication of the model may highlight interesting implications as the uncertainty (or volatility) of innovation processes within technological domains -- with higher variance often associated with emerging or rapidly evolving fields -- has long attracted the broader innovation literature \citep{jalonen2011uncertainty, jalonen2012uncertainty, allen2013complexity}. This shows the potential of our setting, whose stochastic nature makes it possible to characterize not only average success dynamics but also how fluctuations and cross-category comovements evolve over time. \\
Notice that the framework abstracts from exogenous latent common factors. Therefore, the result below should be interpreted as a model-based implication of the interaction structure $\Gamma$, rather than as a claim that all observed comovement in the data must necessarily arise from direct interactions alone. The next theorem is a consequence of~Theorem~\ref{th-P} and it characterizes how volatility and cross-category dependence evolve over time.

\begin{theorem}\label{cor-X} 
Under assumptions {\Asa} and {\Asb}, denote by $\gamma^* \in (0, 1]$ the
Perron-Frobenius eigenvalue of $\Gamma$ and by $\vu = (u_h)_h$ its corresponding (unique up to a multiplicative non-zero constant) left eigenvector. Then, for each $h\in{\mathcal H}$, 
we have 
$$
t^{1-\gamma^*} Var[X_{t+1,h}]\longrightarrow |u_h|\,\alpha(\vu)\,, 
$$ 
where $\alpha(\vu)>0$ (and such that $\alpha(C\vu)=\alpha(\vu)/|C|$ for each constant $C\neq 0$). 
 Moreover, for each pair $h\neq j$ of different processes, we get  
$$
t^{2(1-\gamma^*)}cov(X_{t+1,h}, X_{t+1,j})\longrightarrow u_hu_j\sigma^2(\vu)\,,
$$
where $\sigma^2(\vu)>0$ (and such that $\sigma^2(C\vu)=\sigma^2(\vu)/C^2$ for each constant $C\neq 0$), so that 
$$
t^{1-\gamma^*} \rho(X_{t+1,h},X_{t+1,j})\longrightarrow \sqrt{u_hu_j}\,\tfrac{\sigma^2(\vu)}{\alpha(\vu)}\,.
$$
\end{theorem}

Hence, the ratio of the variances $Var[X_{t,h}]/Var[X_{t,j}]$ converges 
to the ratio $u_h /u_j$ of the relative eigenvector centrality scores (with respect to $\Gamma^\top$) of the two nodes (processes) $h$ and $j$. Moreover, the correlation coefficient between the observations related to any pair of different processes converges to zero at the rate $1/t^{(1-\gamma^*)}$ when $\gamma^*<1$ and converges to the same strictly positive value when $\gamma^*=1$.\\
In short, a balanced system ($\gamma^* = 1$) supports persistent uncertainty and correlated success across technological categories. By contrast, an unbalanced system ($\gamma^* < 1$) leads over time to more asymmetric and progressively decoupled innovation dynamics. In other words, when interaction is strong enough to keep the system balanced, categories continue to display common fluctuations and persistent comovement; when reinforcement is subcritical, cross-category correlations vanish asymptotically and success dynamics become increasingly separated across categories.\\
 The interpretation that follows, moreover, is consistent with the literature emphasizing that innovation outcomes depend on technological proximity and network linkages rather than evolving in isolation \citep{jaffe1989characterizing, taalbi2020evolution}.\\

\indent Finally, in order to extend the analysis from category-level success counts to source-target success counts, we develop Theorem~\ref{th-S-2}. As explained in the following, this theorem examines not only how many successful patents there are in each target category $h$, but also traces the source category $k$ to which each successful patent originally belonged. Namely, in our particular context, we can enrich the model assuming that the number $S_{t,k,h}$ of successes until time-step $t$ in category (process) $h$ coming from belonging category $k$ is of the form $S_{t,k,h}=\sum_{n=1}^t X_{n,h} Y_{n,k}$, where $Y_{n,k}$ takes value $1$ if the category of patent $n$ is $k$ and $0$ otherwise and  
\begin{itemize}
\item[{\Asc}] $\vY_n=(Y_{n,k})^\top_{k\in{\mathcal H}}$ is independent of $\vX_n=(X_{n,h})^\top_{ h\in {\mathcal H} }$ and of all the past until time-step $n-1$ with $
P(Y_{n,k}=1)=\pi_k\in (0,1)$ (where $\sum_{k\in{\mathcal H}} \pi_k =1$).  
\end{itemize}

Assumption {\Asc} should be interpreted as a benchmark introduced for tractability. Its purpose is to isolate the role of the composition of patent activity across source categories, rather than to provide a full empirical description of source-category assignment. In particular, the assumption abstracts from the possibility that patents sort endogenously across source categories and from the possibility that source-category composition co-evolves with success dynamics.\\
Under these assumptions, we obtain the following result.

\begin{theorem}\label{th-S-2}
    Under assumptions {\Asa}, {\Asb} and {\Asc}, denote by $\gamma^* \in (0, 1]$ the
Perron-Frobenius eigenvalue of $\Gamma$. Then, for each pair $h,k\in{\mathcal H}$, we have 
   $$
   \frac{S_{t,k,h}}{t^{\gamma^*}}\stackrel{a.s.}\longrightarrow \pi_k S_{\infty, h}
   $$
   and so  $S_{t,k,h}/S_{t,j,h}\stackrel{a.s.}\longrightarrow \pi_k/\pi_j$. 
\end{theorem}

Theorem~\ref{th-S-2} provides a formal representation of how the composition of innovation inputs shapes the distribution of successful outcomes across technological domains. Specifically -- under the benchmark assumption \Asc-- it shows that, asymptotically, the share of successful innovations in a target category reflects the origin distribution of patent activity. This result helps clarify how early category composition can propagate into long-run patterns of successful innovation across interacting technological domains.

The remaining material in this section provides visual illustrations of two implications of the framework that are especially relevant for the innovation context studied in the paper. First, because categories interact through $\Gamma$, localized increases in success can propagate across related categories. Second, because the framework is stochastic, it yields predictions not only for cumulative success dynamics but also for the evolution of volatility and cross-category comovement. These simulations help make the dynamic logic of the model more transparent before turning to the patent data.
%%%%%%%%%%%%%%%%%%%%%%%%%%%%%%%%%%%%%%%%%%%%%%%%%%%%%%%%%%%%%%%%%%%%%%%%%%%%%%%%

\subsection{Model implication: simulations with a shock}\label{sec:shock}
\begin{figure}[tbhp]
\centering
\includegraphics[height=8cm]{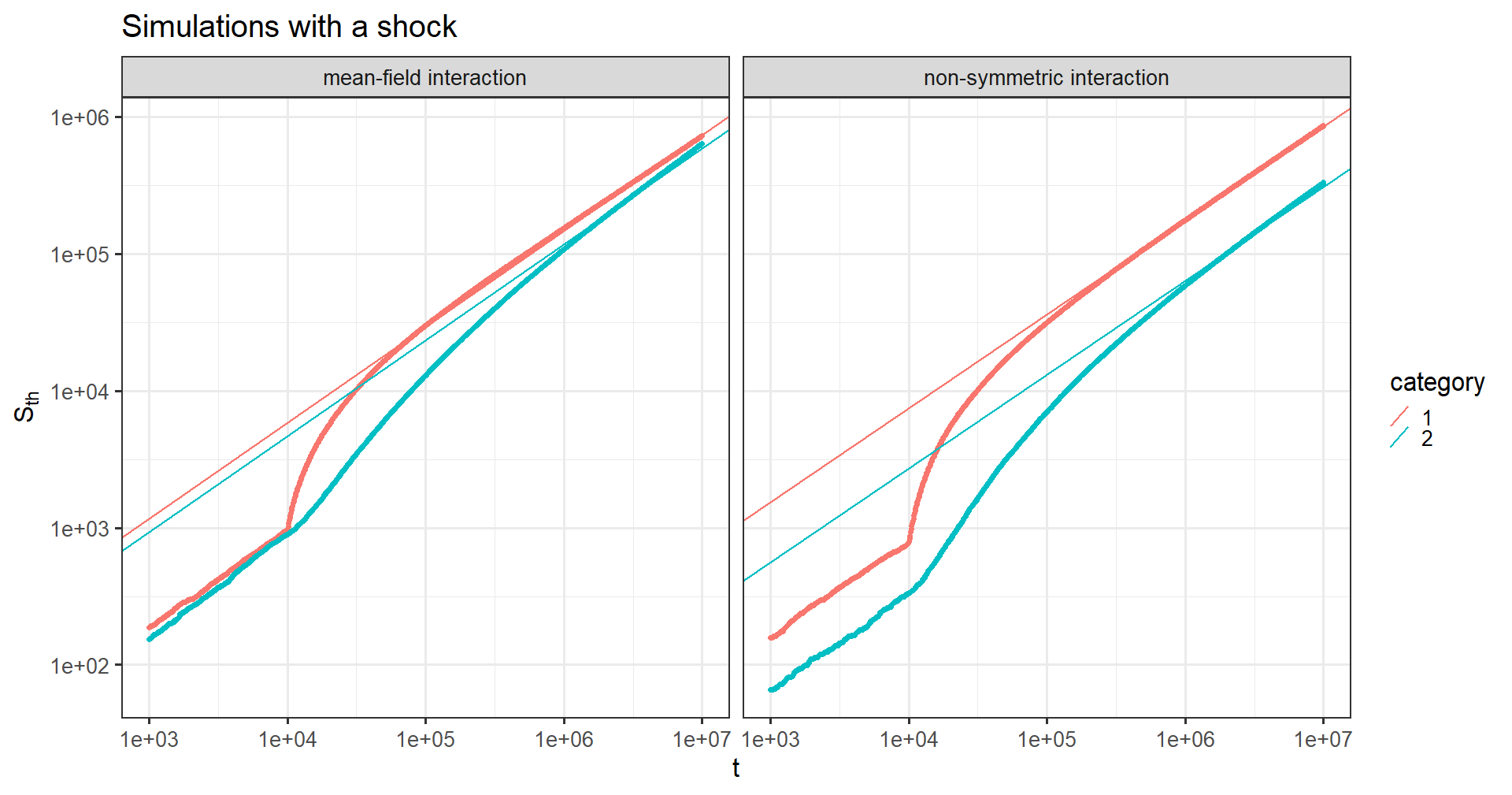}  
\caption{Linear behavior of the successes counter $S_{t,1}$ (red line) and $S_{t,2}$ (blue line) along $t$, in $\log_{10}-\log_{10}$ scale, in two different scenarios with a positive shock on process $1$. 
The two scenarios are distinguished by the type of interaction, i.e. by the form of the interaction matrix $\Gamma$ used in the simulations. The shock on process $1$ occurs at time $t_{shock}=10^4$ and the system is observed until time-step $t=10^7$. We can observe the sudden rise of $S_{t,1}$ (red line) caused by the shock and how the positive effect of the shock on process $1$ propagates to process $2$ (consequent progressive rise of the blue line). We can also note that the slope of the two lines and the distance between their intercepts are the same before and after the shock, because the interaction matrix $\Gamma$ does not change: indeed, the slope is equal to $\gamma^*$ and the distance between the intercepts corresponds to the quantity $\log_{10}(u_1/u_2)=\log_{10}(u_1)-\log_{10}(u_2)$.
%
%The lines are obtained by a least square
%interpolation using a subsample of size $10^4$ of the simulated data imposing a common slope equal to the true value $\gamma^*$ derived from the matrix $\Gamma$ of the corresponding scenario.
}
\label{fig:simulation_with_shock}  
\end{figure}
It is worthwhile to illustrate what the model implies for propagation conditional on the cross-category interaction structure $\Gamma$. If innovation success is cumulative and interdependent, then a shock affecting one category should not remain confined to that category, but should transmit to other connected categories through the inter-dependent reinforcement mechanism. The simulation below visualizes this implication of the model showing how the effect of a \emph{shock} to one process propagates to the others through the interaction structure. A model that accounts for the propagation of innovation shocks, moreover, is useful because such spillovers are frequently observed in practice \footnote{A vivid example of this phenomenon is the surge in pharmaceutical innovation during the COVID-19 pandemic, which triggered subsequent innovations in other fields such as logistics \citep{dovbischuk2022innovation} and IT \citep{li2022digital}, and also prompted a wave of open innovation practices through which inventors widely disseminated their contributions \citep{brem2021implications, lee2021convergence,  ho2023research}.}.

\indent We consider here a system with $N=2$ processes starting with the same initial composition, $\theta_h=1/2$ and $c_h=1$. The purpose is to analyze two simple benchmark interaction structures. The first is a symmetric benchmark, in which the two categories are equally connected and spillovers are homogeneous across them. The second is an asymmetric benchmark, in which cross-category effects are directional and one category is more central in the interaction structure than the other. In this way, the simulation illustrates how shock propagation depends not only on the presence of interdependence, but also on its structure. We take into account the following two scenarios:
\begin{itemize}
\item mean-field interaction, that is a symmetric interaction matrix $\Gamma$ of the form
\begin{equation}\label{eq:elements_Gamma_mean-field}
\gamma_{j,h}=
\begin{cases}
\gamma^{*}\left(\iota/N+(1-\iota)\right)\quad &\mbox{if } j=h \\
\gamma^{*}\,\iota/N\quad &\mbox{if } j\neq h
\end{cases}
,
 \end{equation}
with $\gamma^*,\,\iota\in (0,1]$ \footnote{We refer to this kind of interaction as the mean-field interaction, because in this case the term $\sum_{j\in\mathcal{H}} \gamma_{j,h} S_{t,j}$ in \eqref{eq-model} becomes $\gamma^*\left(\iota \sum_{j\in{\mathcal H}}S_{t,j}/N +(1-\iota) S_{t,h} \right)$ and so the conditional probability $P_{t,h}$ depends on a convex combination between $S_{t,h}$ and the averaged value of all the $S_{t,j}$ in the system.}, where we have chosen $\gamma^{*}=0.7$ and $\iota=0.9$, so that 
$$
\Gamma=\begin{pmatrix}
0.385 & 0.315\\
0.315 & 0.385
\end{pmatrix}\,;
$$
\item non-symmetric interaction matrix $\Gamma$ with $\gamma^{*}= 0.685$ and $\vu=(1.394, 0.575)^\top$, i.e.
$$
\Gamma=\begin{pmatrix}
0.50 & 0.20\\
0.45 & 0.20
\end{pmatrix}.
$$
\end{itemize}
The \emph{shock} occurs at time $t_{shock}=10^4$ and it acts as follows: 
at time $t_{shock}=10^4$, the probability of having a success for process $1$ is increased 
(that is, we give a positive shock to process $1$) by replacing the parameters $\theta_1=1/2$ and $c_1=1$ by the new ones $\theta_{shock,1}=1/2+10^4$ and $c_{shock,1}=1+10^4$. These new values of the parameters remain in the dynamics of process $1$ for all subsequent time-steps until $t=10^7$. The parameters for process $2$ are unchanged: $\theta_2=1/2$ and $c_2=1$ remain the same throughout. The number of successes for the two processes, namely $S_{t,1}$ and $S_{t,2}$, are then observed. Figure~\ref{fig:simulation_with_shock} shows that the positive shock causes the number of successes for process $1$ to rise and that this effect also propagates to process $2$ through the interaction terms in the dynamics. It also shows that, although the shock shifts the levels of the trajectories, it does not alter their long-run scaling properties, which remain governed by the interaction structure summarized by $\Gamma$.

%%%%%%%%%%%%%%%%%%%%%%%%%%%%%%%%%%%%%%%%%%%%%%%%%%%%%%%%%%%%%%%%

\subsection{Model implication: simulations on variance and covariances} \label{sec:simulations}
\begin{figure}[tbhp]
\centering
\includegraphics[height=8cm]{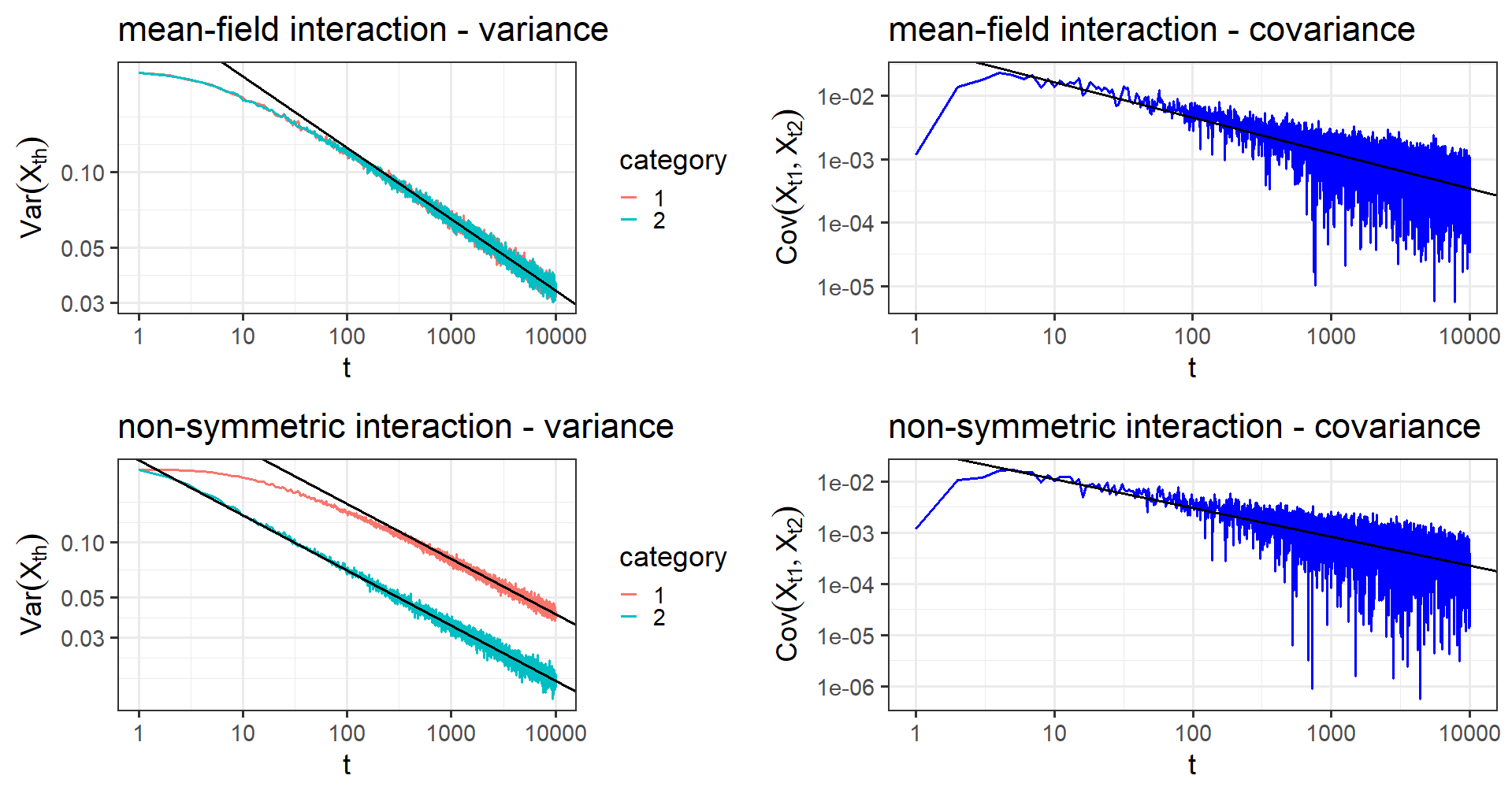}  
\caption{Linear behavior of $Var[X_{t,h}]$ (left panels) and $cov[X_{t,1},X_{t,2}]$ (right panels) along $t$, in $\log_{10}-\log_{10}$ scale, in two different scenarios.
These quantities have been estimated by performing $10^4$ independent simulations and the system is observed until time-step $t=10^4$.
The two scenarios are distinguished by the type of interaction, i.e. by the form of the interaction matrix $\Gamma$ used in the simulation.}
\label{fig:var_cov}  
\end{figure}
{In this section, we illustrate through simulations the implications of Theorem~\ref{cor-X} in the empirically relevant unbalanced case \(\gamma^*<1\). The point is to show that the stochastic predictions of the model are not limited to asymptotic statements: even in finite simulations, the framework implies that fluctuations in category-specific success and cross-category comovement weaken over time when reinforcement is subcritical. In other words, as successful innovations become rarer per patent opportunity, uncertainty at the category level also declines and success dynamics across categories become progressively less correlated. Figure~\ref{fig:var_cov} shows these implications in the same two benchmark interaction structures considered in Subsection~\ref{sec:shock}.}\\
Specifically, we run \(10^4\) independent simulations of \(N = 2\) processes up to a final time of \(t = 10^5\), considering the scenarios of mean-field and non-symmetric interaction outlined in Subsection~\ref{sec:shock}. In Figure~\ref{fig:var_cov}, we observe that the variances of both the processes \((X_{t,1})_t\) and \((X_{t,2})_t\) decays according to a power law with exponent \(1 - \gamma^*\) (left panels). Similarly, the covariance between the processes \((X_{t,1})_t\) and \((X_{t,2})_t\) exhibits a comparable asymptotic behavior, with an exponent of \(2(1 - \gamma^*)\) (right panels).
\\
\indent The plots are presented in a log-log scale, where the power-law decay appears as linear behavior, with the exponent corresponding to the slope of the fitted line obtained from linear regression. In the left panels of Figure~\ref{fig:var_cov}, the difference between the intercepts can be interpreted as the logarithmic ratio between the variances of the two categories, i.e., \( \frac{\text{Var}[X_{t+1,1}]}{\text{Var}[X_{t+1,2}]} \), which converges to \( \frac{u_1}{u_2} \).
\\
\indent In the first scenario of mean-field interaction, the estimated common slope of -0.286 in Figure~\ref{fig:var_cov} (top-left panel) is in excellent agreement with the theoretical value of \(-(1 - \gamma^*) = -0.3\). In the second scenario of nonsymmetric interaction, the estimated common slope (bottom-left panel) is -0.302, which is also very close to the theoretical value of \(-(1 - \gamma^*) = -0.315\). The difference in intercepts of the two processes in the two scenarios is 0.00023 and 0.367, respectively, which aligns well with the theoretical predictions: \(\log_{10}(u_1/u_2) = 0\) for the mean-field case, and \(\log_{10}(u_1/u_2) = \log_{10}(1.394/0.575) = 0.385\) in the nonsymmetric interaction scenario.
\\
\indent For the covariances, Figure~\ref{fig:var_cov} shows that the estimated slope in the mean-field scenario (top-right panel) is -0.559, which is in very close agreement with the theoretical value of \(-2(1 - \gamma^*) = -0.6\). Similarly, in the nonsymmetric interaction scenario, the estimated slope (bottom-right panel) is -0.563, which is in good agreement with the theoretical value of \(-2(1 - \gamma^*) = -0.629\).\\~\\
\indent {Overall, the simulations confirm that, as successful innovations become rarer per patent opportunity, innovation dynamics across categories also become progressively less correlated. This complements the earlier results on sublinear growth and stable long-run shares, and helps clarify the broader implication of the model: cumulative and interdependent innovation need not remain equally volatile or tightly coupled as the system expands.}
%%%%%%%%%%%%%%%%%%%%%%%%%%%%%%%%%%%%%%%%%%%%%%%%%%%%%%%%%%%%%%%%

\section{Data analysis}\label{sec:data_analysis}

%This section examines whether the main aggregate implications derived in Section~\ref{sec:mathbkg} are present in the patent data. In line with the logic of the paper, the empirical analysis is organized around the observable objects that provide the main bridge between the model and the data. We first construct the empirical system of category-specific success indicators and then study whether the patent data display the aggregate regularities emphasized in the Introduction and formalized in Theorems~\ref{th-S}, \ref{cor-X}, and \ref{th-S-2}. In particular, we focus on the cumulative counts of successes, their relative long-run proportions across categories, and their source-target decomposition. We then complement this comparison with a parsimonious inference exercise on interaction intensity under a mean-field restriction.

Starting from the category-specific success indicators defined in Section~\ref{sec:cfc}, we now turn to their empirical evolution in the patent data following the results derived in Section~\ref{sec:mathbkg}. The most direct quantities to examine are the cumulative success counts and their relative behavior across categories, because these are the observable objects through which the model yields its main aggregate implications. We therefore study their scaling properties, their long-run proportions, and their source-target decomposition, and then report a complementary inference exercise on interaction intensity under a mean-field restriction.

Our dataset consists of the GLOBAL PATSTAT database\footnote{The dataset is maintained at the IMT School.}. More specifically, we collected all the granted US patent families (DOCDB) published in the period $[1980-2018]$, with their exact (full) date of publication and their CPC-1 category. 
Moreover, for each patent $n$, we know if it has been cited by subsequent patents (published in the considered period) and which are the citing patents. 

We limit ourselves to the CPC categories $h\in{\mathcal H}=\{A,\, B,\, C,\, D,\, E,\, F ,\, G,\, H\}$ (i.e. we exclude category $Y$ \footnote{This category has been excluded due to its broad scope and ambiguity, which are well-documented in the literature \citep{leydesdorff2017mapping, rainville2025tracking}. Although it is often broadly referred to as the class encompassing green patents, it is widely recognized in the green innovation literature that only the subclasses $Y02$ and $Y04$ accurately represent genuine green technologies \citep{corrocher2021international, barbieri2023regional}.}). 
We take the size $T$ of the time-window for the computation of the index $I_{n,h}$ equal to $5$ years (coherently with other previous studies, e.g. \cite{squicciarini2013measuring}). Therefore, since the available dataset contains patents with publication year before or equal $T_f=2018$, we perform the analysis on  the patents with publication year in the period between $1980$ and $(T_f-T)=2013$, so that, for all of them, we have the entire time-window to compute the index. As explained in Sec.~\ref{sec:cfc}, this is done in order to neutralize any truncation effect in citations, since all the considered patents have the same time-window to receive citations.  
Finally, we fix the threshold $\tau=0.8$ (see Section~\ref{choice_threshold-app} for some details on the choice of the threshold). We thus obtain a matrix with $N=\text{card}({\mathcal H})=8$ columns, where 
each row corresponds to a patent $n$. The patents (and so the rows) are ordered with respect to their publication (full) date. The total number of rows is $n_{tot}=5\,004\,253$. The matrix entry $x_{n,h}$ is equal to $1$ if the patent $n$ is a success for category $h$, i.e. if its value of the index is above the threshold $\tau$, or equal to $0$ if $n$ is not a success. Hence, the constructed matrix can be seen as the realization of a finite system $\{ X^{(h)}=(X_{n,h})_{n} \,:\, h\in {\mathcal H}\}$ of Bernoulli processes. The patent ordering induces the relevant notion of \emph{system time} used throughout the paper, and the remainder of the section asks whether the aggregate behavior of the real data $\{x_{n,h}\}$ is in agreement with the theoretical results of the previous section.

\subsection{Aggregate empirical patterns and their connection with the model}
\label{sec:aggregate_patterns}

\begin{figure}[tbhp]
\centering
\includegraphics[height=8cm]{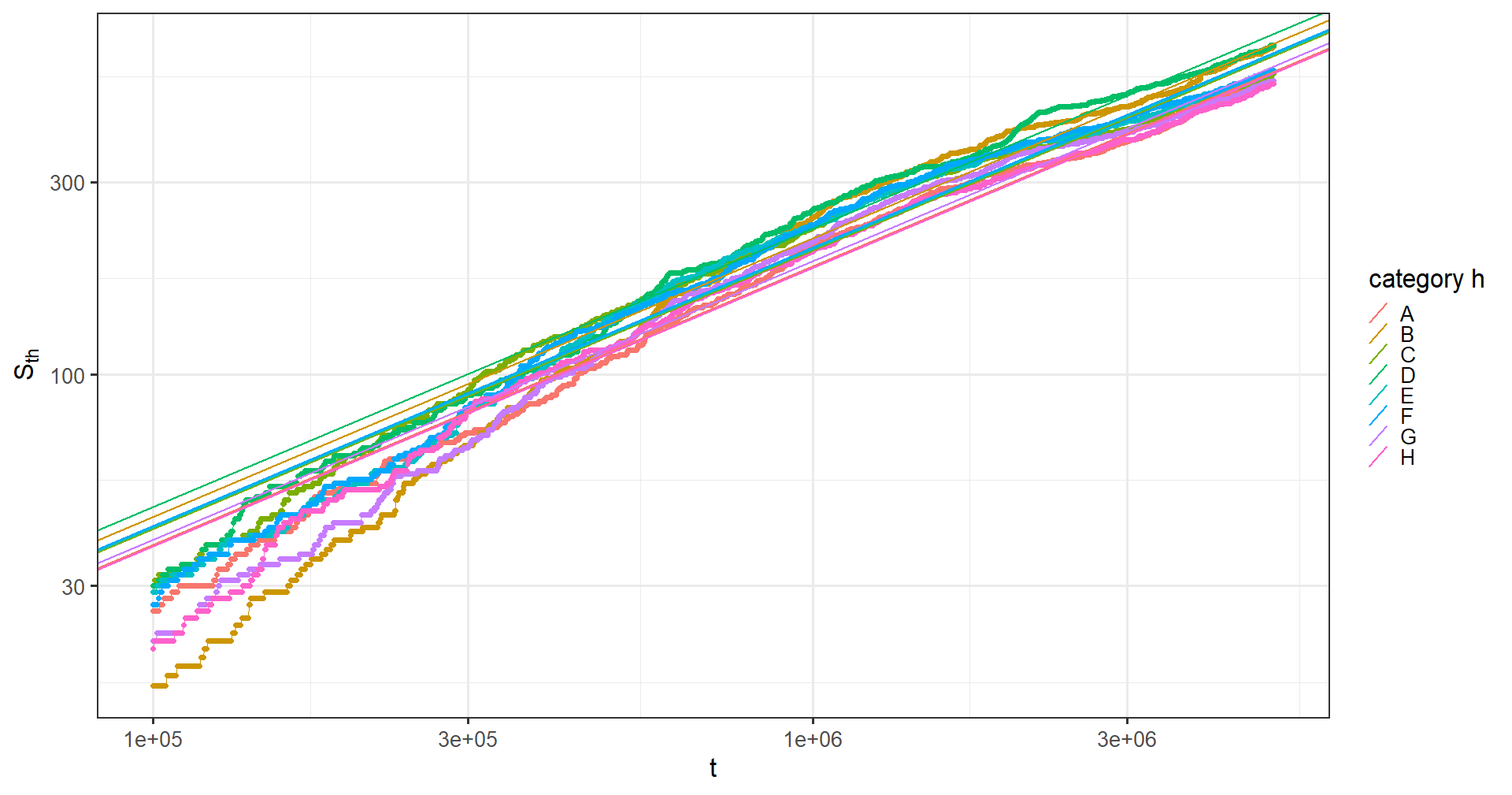}
\caption{Linear behavior of the successes counter $S_{t,h}$ along $t$, in $\log_{10}-\log_{10}$ scale, for each category $h$. The lines are obtained by a least square
interpolation using a subsample of size $10^4$ of the data set. 
We can appreciate how the lines exhibit the same slope, which has been estimated to be equal to $\widehat{\gamma}^*=0.689$.
Indeed, the goodness of fit $R^2$ index obtained imposing a common slope is $0.969$ and so basically equal to the one obtained allowing the slopes to be different across categories, i.e.~$0.972$. The emergence of a common Heaps' exponent across all categories is suggestive of shared interdependence, i.e. of the fact that these innovation processes are not evolving independently but are shaped by shared systemic dynamics. Moreover, the sub-linear growth reflects diminishing returns to cumulative knowledge, increasing innovation complexity, and a bounded scalability in innovation outputs over time. The figure is plotted in system time, so the estimated sublinear scaling should be interpreted as a decline in success per patent opportunity rather than as a mechanical statement about calendar-time trends.}
\label{fig:S}
\end{figure}

We begin from the cumulative success counts, since these are the observable quantities that most directly correspond to the main implications of the model. Theorem~\ref{th-S} predicts that, under the maintained assumptions, the number of cumulative successes in each category should grow with the same Heaps' exponent $\gamma^*$. Figure~\ref{fig:S} examines this first implication in the patent data.

Figure \ref{fig:S} provides the asymptotic behavior, in $\log_{10}-\log_{10}$ scale, of every process $S_{t,h}$, that represents the number of successes observed in category $h$ until time-step (patent) 
$t$. We can appreciate how the lines exhibit the same slope, which indicates that the processes have the same Heaps' exponent. This is exactly in accordance with Theorem \ref{th-S}. The emergence of a common Heaps' exponent across all categories is therefore consistent with the model prediction that category-specific cumulative successes share a common long-run scaling parameter. At the same time, and coherently with the discussion in Section~\ref{sec:mathbkg}, this pattern should be interpreted with some care: it is compatible with the presence of systemic interdependence, although it does not by itself distinguish direct cross-category interaction from co-movement induced by broader common factors.\footnote{This pattern is compatible with the presence of systemic interdependence, although it does not by itself distinguish direct cross-category interaction from co-movement induced by broader common factors.} We provide here, therefore, evidence consistent with the view that innovation success across categories evolves in an interdependent way rather than as a collection of completely separate category-specific processes.

\indent The value of the common Heaps' exponent, 
estimated as the common slope of the lines in the $\log_{10}-\log_{10}$ plot, 
is $\widehat{\gamma}^*=0.689$.\footnote{In order to measure the uncertainty of the slopes estimated for each category $h$ at a given time-step, we have performed a linear regression with random effects, on both intercept and slope and using a subsample of size $10^4$ of the data set. 
The estimated values of the slopes, one for each category $h$, presents a mean equal to 
$\widehat{\gamma}^*=0.689$ and standard deviation equal to $0.044$.} 
The estimate $\widehat{\gamma}^*=0.689$ therefore implies sublinear growth of cumulative successes in system time. Interpreted through the model, this means that the probability that the next observed patent is a success declines as the number of patents already observed increases. Put differently, successful patents continue to occur, but they become less frequent relative to the expanding set of patent opportunities. This interpretation refers to system time rather than calendar time: Figure~\ref{fig:S} does not imply that yearly success rates must decline for purely mechanical reasons, but that cumulative success grows less than proportionally with the scale of inventive activity. This is the economically relevant sense in which success becomes harder to obtain as the system expands. More broadly, the common sublinear pattern across categories points to a form of bounded scalability in innovation outputs: innovation continues, but the marginal probability of producing an additional success in any category diminishes, consistently with the literature on declining research productivity and increasing knowledge burdens \citep{bloom2020ideas,jones2009burden}.

\begin{figure}[tbhp]
\centering
\includegraphics[height=8cm]{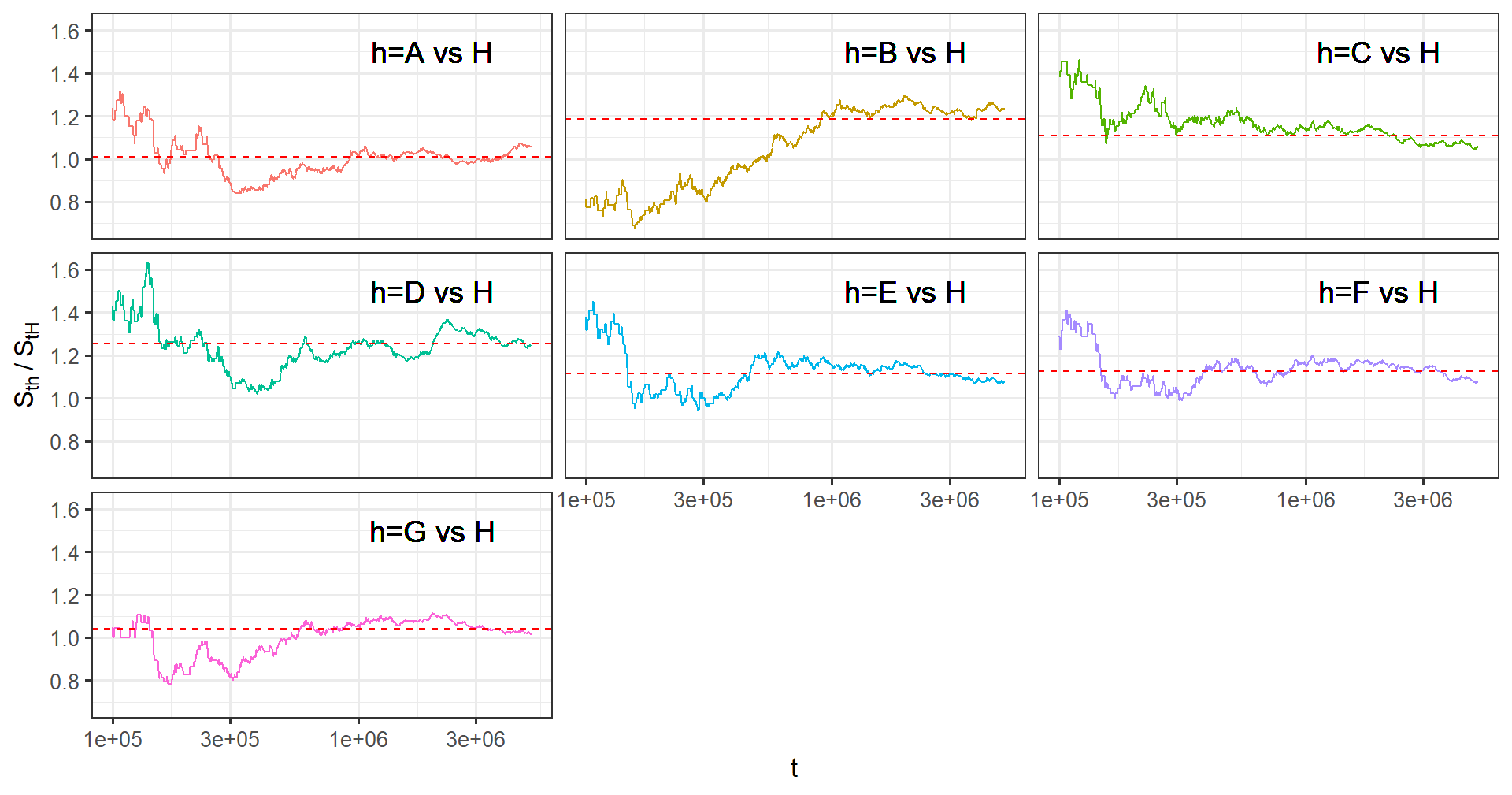}  
\caption{Plot of the process $S_{t,h}/S_{t,H}$ along time $t$, for all the categories $h\neq H$. Category $H$ is arbitrarily chosen to be the baseline category. The horizontal dashed red lines  
represent an estimation of the ratio $u_h/u_H$, obtained as 
$10^{d_{h,H}}$, where $d_{h,H}$ is the difference between the intercepts of the regression lines in Figure \ref{fig:S} for the categories $h$ and $H$.  This long-run convergence across categories highlights a structural stabilization in innovation outcomes, despite increasing complexity and declining marginal returns to knowledge production. Moreover, note that all the estimated ratios $u_h/u_H$ are around $1$. This means that in the long run the numbers of successes in the different categories tend to coincide.}
\label{fig:log_ratio_S}  
\end{figure}

The second empirical implication emphasized in the Introduction and derived in Theorem~\ref{th-S} concerns not the common growth exponent itself, but the long-run relative proportions of successes across categories. If cumulative success counts scale with the same exponent, then their ratios should stabilize over time. Figure~\ref{fig:log_ratio_S} examines this implication.

Figure \ref{fig:log_ratio_S} shows, for each category $h$, the convergence of the process $(S_{t,h}/S_{t,H})_t$ toward the quantity $u_h/u_H$, 
estimated as $10^{d_{h,H}}$ where $d_{h,H}$ is the difference between the intercepts of the regression lines in Figure \ref{fig:S} for the two processes related to the pair $(h,H)$ of categories.\footnote{Category H is arbitrarily chosen to be the baseline category. Any other category can be used as a baseline.} Also this fact is in accordance with Theorem \ref{th-S} and it means that the relative number of successes across categories stabilizes over time. In other words, although categories may differ in levels, their long-run relative success shares tend to settle around stable proportions. This interpretation is in line with work emphasizing that technological interdependencies shape persistent patterns of innovation across domains \citep{taalbi2020evolution}, even when producing additional influential innovations becomes progressively harder over time \citep{bloom2020ideas}. Moreover, the limit corresponds to the ratio of the respective components of the eigenvector-centrality scores $\vu$ (with respect to $\Gamma^\top$). \\

Taken together, Figures~\ref{fig:S} and \ref{fig:log_ratio_S} provide the empirical counterpart of the two main aggregate implications highlighted in the Introduction: cumulative successes grow sublinearly in system time and their relative shares across categories stabilize in the long run. For this reason, they constitute the core empirical evidence of the paper.

We now consider an extension that tracks successful patents jointly by target category and source category. This part of the analysis is included because Theorem~\ref{th-S-2} refines the aggregate cumulative-success dynamics by showing how they decompose according to the source category from which successful patents originate. In this sense, the following evidence is not a separate empirical claim, but a finer decomposition of the aggregate patterns already documented above.

\begin{figure}[tbhp]
\centering
\includegraphics[height=8cm]{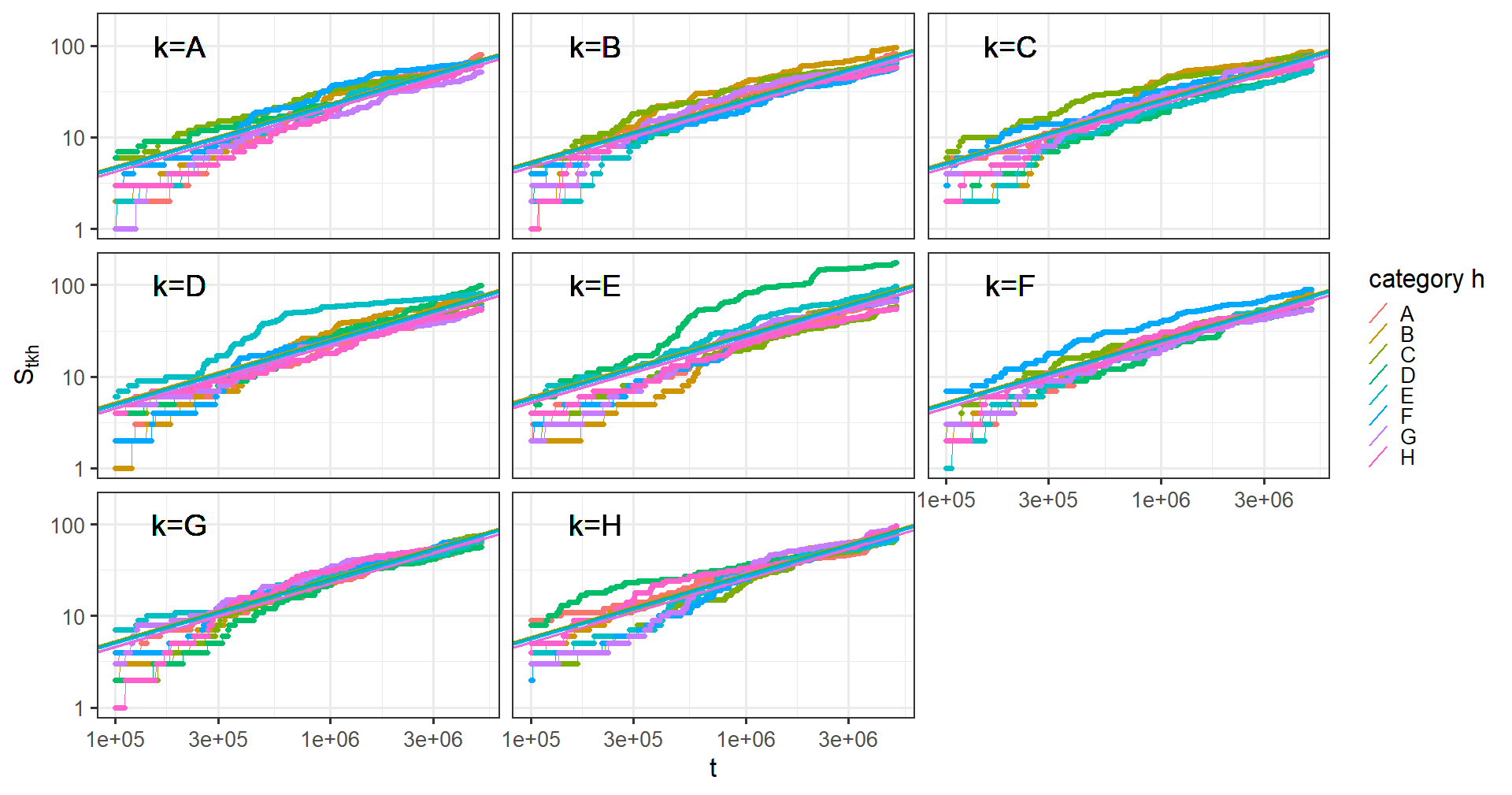}
\caption{Linear behavior of  the successes counter $S_{t,k,h}$ along $t$ in $\log_{10}-\log_{10}$ scale, for each pair $(k,h)$ of categories ($k=$sub-figure, $h=$color).
The lines are obtained by a least square interpolation based on a subsample of the dataset with a slope equal to the previously estimated value $\widehat{\gamma}^*=0.689$. The goodness of fit $R^2$ index is $0.851$,
which is basically the same as the one obtained allowing the slopes to be different across the pairs $(k,h)$ of categories, i.e. $0.855$. The common slope suggests one again the presence of interaction across the categories. Moreover, the sub-linear growth reflects a consistent sub-linear scaling of cross-domain innovation flows, suggesting that while knowledge recombination persists, its marginal productivity diminishes over time across all category pairs.}
\label{fig:S_hk}
\end{figure}

\indent Figure \ref{fig:S_hk} provides the asymptotic behavior, in $\log_{10}-\log_{10}$ scale, of every process $S_{t,k,h}$, that represents the number of successes for category $h$ coming from category $k$ observed until time-step (patent) $t$. We can appreciate how the lines exhibit the same slope, equal to the previously estimated value $\widehat{\gamma}^*=0.689$ in accordance with Theorem \ref{th-S-2}. In this sense, the source-target decomposition inherits the same aggregate scaling structure documented at the category level.

\begin{figure}[tbhp]
\centering
\includegraphics[height=8cm]{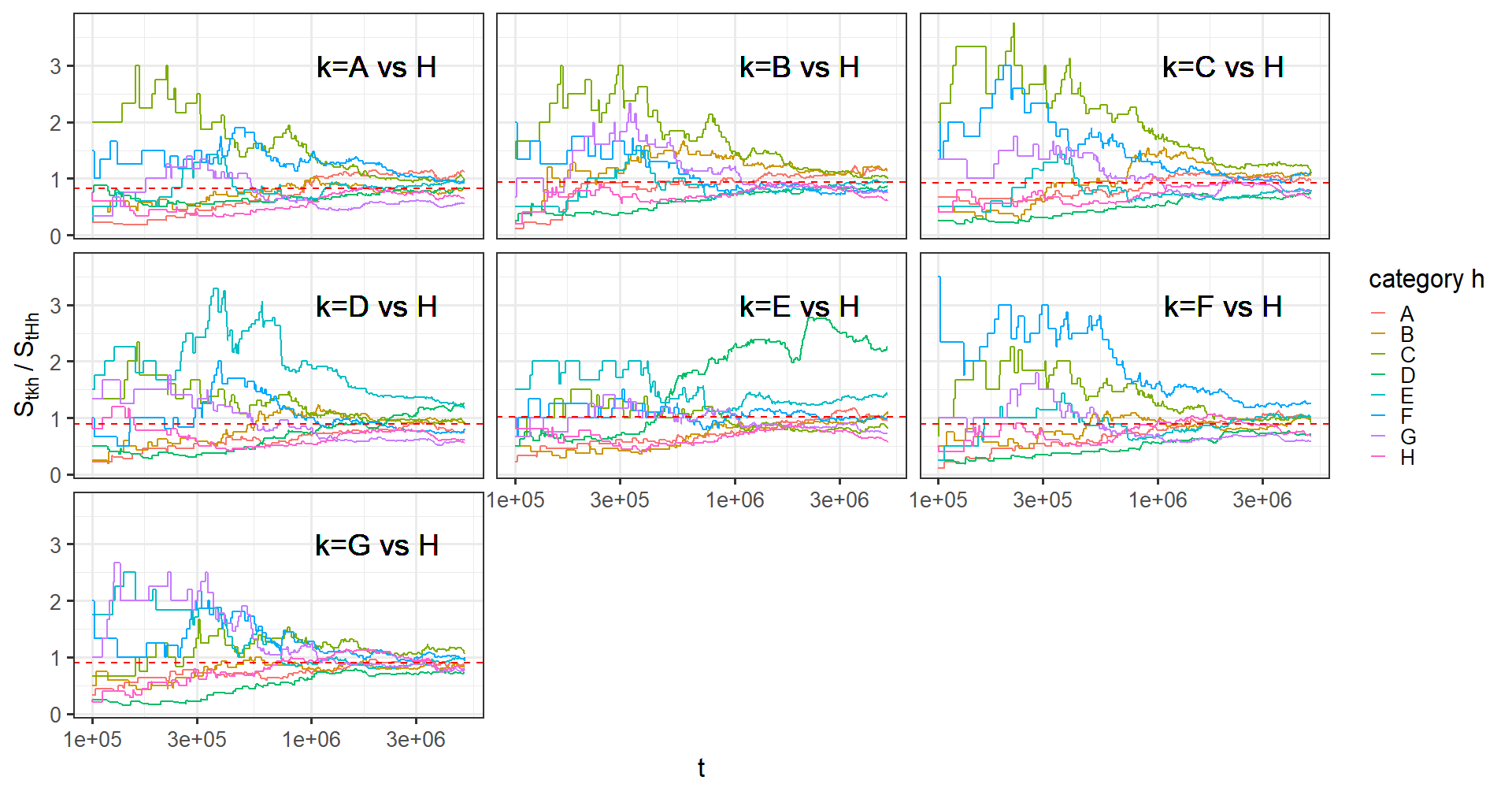}  
\caption{Plot of the process $S_{t,k,h}/S_{t,H,h}$ along time $t$, for all the pairs $(k,h)$ of categories  ($k=$sub-figure, $h=$color), with $k\neq H$. Category $H$ is arbitrarily chosen to be the baseline category. The horizontal dashed red lines  
represent the value $10$ to the power of the differences between the intercepts of the regression lines in Figure \ref{fig:S_hk} for the pairs of categories $(k,h)$ and $(H,h)$. This long-run proportionality links early patenting intensity to long-term innovation success, highlighting the path-dependent and self-reinforcing nature of technological trajectories.}
\label{fig:log_ratio_S_k}  
\end{figure}

\indent Finally, each panel of Figure \ref{fig:log_ratio_S_k} refers to the successes observed in a given category $h$ and shows, for each coming category $k$, the convergence of the process $S_{t,k,h}/S_{t,H,h}$ toward the quantity $\pi_k/\pi_H$, 
estimated as 10 to the power of the difference between the intercepts of the corresponding regression lines in Figure \ref{fig:S_hk}.\footnote{Category H in Figure \ref{fig:log_ratio_S_k} is arbitrarily chosen to be the baseline category. Any other category can be used as a baseline.} Also this fact is coherent with Theorem \ref{th-S-2}. In the long run, therefore, the number of successful innovations in each category ends up matching how much patent activity started in each category. Hence, if a certain type of technology had a greater patent activity early on, it will also tend to have more successful innovations over time\footnote{These patterns are consistent with the benchmark implication in Theorem~\ref{th-S-2}, although that result relies on the stylized assumption \Asc and should therefore be read as an illustrative benchmark rather than as a full structural account of source-category assignment.}.

Although the main empirical analysis of this section focuses on cumulative counts and long-run proportions, it is useful to recall that Section~\ref{sec:mathbkg} also derived an additional implication concerning fluctuations and cross-category correlations in the empirically relevant unbalanced case $\gamma^*<1$. In the theoretical development, this result serves to show that the stochastic framework yields implications not only for average success dynamics but also for the way uncertainty and comovement evolve over time. In the empirical logic of the paper, this implication complements the two core aggregate patterns discussed above by clarifying why sublinear cumulative growth and stable long-run shares do not imply persistent synchronization in category-specific success dynamics.

\subsection{Statistical inference on the interaction intensity under the mean-field assumption}
\label{stat-inference}

The evidence reported so far compares the patent data with the aggregate implications of the model under a general interaction structure. We now add a complementary exercise under a more restrictive specification. Namely, assuming the system is subjected to a mean-field interaction, we can summarize cross-category dependence by a single average interaction parameter and perform a tractable benchmark inference exercise on its magnitude. This part of the section is included for a different purpose than the figures above. The previous evidence asks whether the aggregate patterns predicted by the framework are present in the data; the present exercise asks whether the observed category-specific success counts move similarly enough to be compatible with a given level of average interaction intensity under a symmetry restriction.

A standard approach in innovation studies is to construct measures of technological proximity or network relatedness and then use them in regression or forecasting exercises to assess their effect on innovative performance \citep{jaffe1989characterizing, pichler2020technological, taalbi2020evolution}. By contrast, under the mean-field specification, our test is model-based: it assesses whether the observed category-specific success counts move similarly enough to be compatible with a given level of average interaction intensity. In this sense, the contribution is not to show that technological interdependence matters --- which is well established in the literature --- but to provide an inferential procedure that is structurally tied to the interacting reinforced Bernoulli framework.\\
Namely, assuming the system is subjected to a mean-field interaction (note that in this case we have $u_h/u_j=1$ for each $h,j\in{\mathcal H}$), i.e. 
assuming the interaction matrix $\Gamma$ to be of the form~\eqref{eq:elements_Gamma_mean-field}, 
we can perform a suitable test on the parameter $\iota$, that rules the intensity of the interaction. This inferential exercise should be interpreted as a parsimonious benchmark under a symmetry restriction on the interaction structure. Under the mean-field specification, heterogeneous pairwise technological linkages are summarized by a single average interaction parameter, rather than being allowed to vary across category pairs. The restriction is useful for tractable model-based inference, but it is not intended as a full description of the heterogeneous and directional interdependencies emphasized in the innovation literature \citep{jaffe1986technological, breschi2003knowledge, colladon2025new}
\\ Notice that, under the mean-field specification, the parameter $\iota$ measures how strongly categories are tied together through a common interaction component. The test checks whether the category-specific success counts move similarly enough to be compatible with a given level of average interaction across categories.\\
For a two-sided test with
 $H_0:\;\iota=\iota_0$, where $\iota_0\leq 1$ and $\iota_0 >1/2$  
(the required condition in the theoretical result for having the 
  second eigenvalue of $\Gamma$ strictly smaller than $\gamma^*/2$),
  we can use
  the test statistic (see Appendix~\ref{app-proofs} for the technical details) 
  \begin{equation}\label{eq:test_statistics_mean_field_case_Gamma}
2\Delta_0\frac{\big\|\vS_t-\widetilde{S}_t\vone
  \big\|^2}{\widetilde{S}_t}  
\stackrel{d}\sim \chi^2(N-1)\quad\mbox{under } H_0,
\end{equation}
where $\Delta_0=\iota_0-\frac{1}{2}$ 
and $\widetilde{S}_t=\tfrac{\vone^\top\vS_t}{N}=\sum_{h\in{\mathcal H}} S_{t,h}/N$.
Since this statistic is increasing in $\iota_0$, it works well also
for the one-sided test with $H_0:\;\iota\geq \iota_0$, 
  with $\iota_0\in (1/2,1]$.
  Applying this one-sided test
 to our real dataset of patents,  
 we obtain different
 $p$-values, one for each tested value $\iota_{0}$,
 that are collected in Table~\ref{table:Table-pvalues}.
We can see that, if we choose the significance level $\alpha=0.05$, then 
the minimum value of $\iota_0$ at which we can 
reject the null hypothesis is 
$\iota_{0}=0.75$. \\

Taking the value of the parameter $\gamma^*$ equal to the estimated one, i.e.~$\widehat{\gamma}^*=0.689$, we have also computed an estimate of the parameter of interest $\iota$ maximizing the 
(joint) likelihood of the processes $X^{(h)}=(X_{n,h})_n$, $h\in{\mathcal H}$, i.e.
$$L(\iota\,|\, x_{n,h},\,h\in{\mathcal H},\,n=1,\dots,n_{tot})=\prod_{n=1}^{n_{tot}}\prod_{h\in {\mathcal H}} P_{h,n}^{x_{h,n}}(1-P_{h,n})^{1-x_{h,n}}.$$ 
The obtained estimate is in accordance with the results of Table~\ref{table:Table-pvalues}. Indeed, the maximum-likelihood estimate of the interaction intensity $\iota$ is $\widehat{\iota}=0.643$ and from Table~\ref{table:Table-pvalues} we get that, at the level $\alpha=0.05$, we can reject the null hypothesis $H_0:\iota\geq 0.75$, while we do not have enough statistical evidence to reject for smaller values of $\iota_0$.

\begin{table}[htbp]  
  \centering
\begin{tabular}{||c|rrrrrrrrr||}
\hline
\hline
$\iota_{0}$ & 0.55 & 0.6 & 0.65 & 0.7 & 0.75 & 0.8 & 0.85 & 0.9 & 0.95 \\ 
p-value & 0.893 & 0.562 & 0.273 & 0.113 & 0.042 & 0.015 & 0.005 & 0.002 & $<0.001$ \\
\hline
\hline
\end{tabular}
\caption{$p$-values associated to the hypothesis test with 
      $H_{0}: \iota\geq \iota_{0}$ under the mean-field interaction .}
 \label{table:Table-pvalues}
\end{table}

The estimated value $\hat{\iota}=0.643$ points to a positive but non-maximal degree of average interaction across technological categories. Read together with the hypothesis-test results in Table~\ref{table:Table-pvalues}, this suggests that the patent data are not compatible with a very high degree of homogeneous interaction across categories, while they are clearly inconsistent with the view that categories evolve independently. In substantive terms, the evidence points to an intermediate configuration: cross-category spillovers are strong enough to generate co-movement in innovative success, but not so strong as to eliminate category-specific dynamics. This interpretation is consistent with the innovation literature showing that technological interdependence and relatedness matter for innovative performance, while substantial heterogeneity across domains remains an important empirical feature \citep{jaffe1989characterizing, taalbi2020evolution, pichler2020technological}. Importantly, our contribution is not to compare $\hat{\iota}$ to a standard coefficient already used in that literature, but to provide a model-based estimate of interaction intensity that is structurally tied to the interacting reinforced Bernoulli framework.

%%%%%%%%%%%%%%%%%%%%%%%%%%%%%%%%%%%%%%%%%%%%%%%%%%%%%%%%%%

\section{Conclusions}\label{concl}
This paper has asked whether innovation success across technological categories can be represented within a single stochastic framework that jointly captures cumulativeness within categories and spillovers across categories. To address this question, we have proposed a model of interacting reinforced Bernoulli processes in which the probability that the next observed patent is a success in category $h$ depends on past successes both within that category and across the others.

The contribution of the paper is to bring together, within one (novel) parsimonious dynamic framework, cumulative reinforcement, cross-category interdependence, and aggregate regularities in innovation outcomes. The value added of the framework, in particular, is that it derives, within a single stochastic structure, joint implications for success probabilities, cumulative successes, relative success shares, and cross-category dependence.

On the empirical side, using granted US patent families from GLOBAL PATSTAT classified at the CPC-1 level, we have constructed category-specific success indicators based on a cohort-normalized forward-citation measure. The empirical analysis shows that the patent data are consistent with the two main aggregate implications emphasized in the paper: cumulative successes grow sublinearly in system time, and their relative shares across categories stabilize in the long run. We have also shown, under the benchmark source-target extension, that the same aggregate scaling logic carries over to the decomposition of successful patents by source category. In addition, under a mean-field restriction, we provide a model-based inference exercise indicating a positive but non-maximal degree of average interaction across categories.

These findings support the paper's core view of innovation success as cumulative and interdependent, while still allowing for category-specific dynamics. They also reinforce an important interpretive point of the analysis: the sublinear result is formulated in \emph{system time}, so it should be read as success becoming rarer per patent opportunity as the system expands, not as a mechanical claim about calendar-time trends. More broadly, the stochastic structure of the framework makes it possible to study not only average success trajectories but also fluctuations, comovement, and shock propagation.

The analysis remains subject to limitations. The interaction matrix is treated as reduced-form and is not itself endogenized; the source-target result is derived under a stylized benchmark assumption introduced for tractability; and the inferential exercise under the mean-field specification is intentionally restrictive, since it summarizes heterogeneous interactions through a single average parameter. Moreover, while the framework abstracts from exogenous latent common factors, some observed co-movement in the data could in principle also reflect broader common influences rather than direct interaction alone.

These limitations also point to directions for further research, including inference for richer interaction structures, more direct empirical testing of the stochastic implications for fluctuations and cross-category comovement, and applications based on alternative taxonomies or alternative indicators of innovative influence.

Overall, the paper provides a parsimonious and empirically interpretable framework for studying how successful innovations accumulate across technological categories in the presence of reinforcement and interdependence. By linking local success dynamics to observable aggregate regularities in patent data, it contributes to the broader effort to formalize innovation as a cumulative, interdependent, and stochastic process.

\clearpage
\newpage
\appendix
% New -- cf. https://tex.stackexchange.com/a/248707
\renewcommand{\thesection}{\Alph{section}} % corrected redefinition of '\thesection'
\makeatletter
\renewcommand\@seccntformat[1]{\appendixname\ \csname the#1\endcsname.\hspace{0.5em}}
\makeatother

%%%%%%%%%%%%%%%%%%%%%%%%%%%%%%%%%%%%%%%%%%%%%%%%%%%%%
%%%%%%%%%%%%%%%%%%%%%%%%%%%%%%%%%%%%%%%%%%%%%%%%%%%%%

 \section{Definitions of novelty in the li\-te\-ra\-tu\-re}\label{appx:litindx}

In patent law, novelty is a legal requirement: for a patent to be granted, the invention must be new relative to the prior art at the time of filing \citep{mueller2024patent,bekkers2020impact}. By contrast, the innovation literature often uses the term ``novelty'' more broadly to refer to concepts such as radicalness, unusual recombination, technological distance, distinctiveness, or ex post technological impact \citep{abbas2014literature,fleming2001recombinant,dahlin2005when, verhoeven2016measuring}. In this appendix, we use ``novelty'' in that broader academic sense unless otherwise stated. The question is therefore not whether a patent is legally novel, but which observable features the literature uses to characterize patents as more or less novel in an economic and technological sense. Table \ref{tab:innovation indices} provides a selective (non-exhaustive) overview of representative approaches, including citation-based, recombination-based, text-based, and composite measures.
\begin{table}[ptbh]
\hspace{-1.5cm}
\scalebox{0.5}{
\begin{tabular}{|l|l|l|l|}
\hline
\textbf{Source} 
& 
\textbf{Novelty Index}                 
& 
\textbf{Quick explanation} 
&   \textbf{Based on}  
\\ \hline
    \citet{squicciarini2013measuring}              
& 
    $\text{Originality}_p = 1- \sum_{j}^{n_p}s_{pj}^2$ 
    & 
\makecell{\footnotesize $s_{pj}$ is the percentage of citations made by 
patent p to patent class $j$ out of the\\ $n_p$ IPC 4-digit (or 7-digit) patent codes contained in the patents cited by patent $p$. \\
Citation measures are built on EPO patents and account for patent equivalents.} 
& 
Citations\\ \hline
     \citet{squicciarini2013measuring}             
     & 
$\text{Radicalness}_p = \sum_{j}^{n_p} \frac{\text{CT}_j}{n_p} \, ; \, \text{IPC}_{pj} \neq \text{IPC}_p$
    & 
\makecell{\footnotesize $CT_j$ denotes the count of IPC-4 digit codes $CPC_{pj}$ of patent $j$ cited in patent $p$ that is\\  not allocated to patent $p$, out of $n$ CPC classes in the backward citations counted at\\ the most disaggregated level available (up  to the 5\textsuperscript{th} hierarchical level). The higher \\ the ratio, the more diversified the array of technologies on which the patent relies upon.
}
&    
\makecell{Citations\\Tech.Knowledge}            
\\ \hline 
\makecell{Own elaboration based \\ on intuitions in \citet{hall2000market} \\
and \\ \citet{lanjouw2004patent}} 
&
\makecell{%
$
\text{ANPCI}_{i,t} = \frac{C_{i,t}}{\lambda_t \cdot \left( \frac{1}{n_t} \sum_{j=1}^{n_t} C_{j,t} \right) }
$
\\
where:
- \( C_{i,t} \) is the number of citations \\ received by patent \( i \) in year \( t \),\\
- \( n_t \) is the number of patents \\filed in year \( t \),\\
- \( \frac{1}{n_t} \sum_{j=1}^{n_t} C_{j,t} \) is the\\ average number of citations per patent in year \( t \),\\
-\( \lambda_t \) is a correction factor to account for\\ citation inflation, which adjusts the \\ average citation counts for inter-year comparability. \\Is the average number of citations per patent across \\ all fields in year $t$, divided by a baseline year (e.g., $t_0$)}
 & 
 \makecell{Intra-year comparison: index is normalized by the average \\ citations within the same cohort (year and field).\\
Inter-year comparison: adjusting for citation inflation $\lambda_t$ with\\ 
 , we should mitigate the bias of comparing\\ older patents with more citations to newer \\ones that haven’t had as much time to accumulate citations.}
 & Citations\\ \hline
% & & & \\ \hline
% & & & \\ \hline
  \makecell{\citet{trajtenberg1997university}\\
  (Modified better version)}  
  & 
  $G_X = 1 - \sum_{j=1}^{M_i} \left( \frac{1}{N} \sum_{i=1}^{N} \frac{T_{ji}^n}{T_{i}^n} \right)^2$ 
  &
  \makecell{\footnotesize $X$ is the focal patent with $Y_i$ patents citing the focal patent $X$, with $i=1,\cdot, N$.\\$T_i^n$ is the total number of IPC $n$-digit classes in $y_i$ 
  \\
$T_{ji}^n$ is the total number of IPC $n$-digit classes in the $j^{th}$ IPC4 digit class in $y_i$ \\
and $j = 1 \dots M_i$ is the cardinal of all IPC4-digit classes in $y_i$} 
& 
\makecell{Citations\\Tech.Knowledge} 
\\ \hline
   Various  & \makecell{Patent quality: \\ Composite indices} & \makecell{\footnotesize Typically based on patent citations, claims, patent renewals and patent family size. \\ Usually different compositions as follows: \textbf{[i]}Patent quality index 4 – 4 components: \\number of forward citations (up to 5 years after publication); patent family size;\\ number of claims; and the patent generality index. Only granted patents are covered \\by the index.
\textbf{[ii]} Patent quality index 4b – 4 components, bis: number of \\forward citations (up to 5 years after
publication); patent family size; corrected claims; and\\ the patent generality index. Only granted
patents are covered by the index. \\ 
\textbf{[iii]} Patent quality index 6 – 6 components: covers the same components\\ as above, plus the number of
backward citations and the grant lag index. \\ \\ \textbf{Entropy based approaches:} Defines a weighting scheme ($W$) for all indicators of a patent (citations, claims...) \\ based on entropy. With $M$ patents, it then constructs $M \cdot W$. A set of negative \\ (patents with the lowest weighted scores across the majority of indicators) and positive patents is \\ then identified and divided; a similarity across good and negative patents is then computed.\\ The following steps are then performed: \textbf{Maximum Similarity}: For each patent in \\ \( M^R \), the maximum similarity to any patent in \( M^N \) is determined. If the maximum similarity \( S(M_i^R, M^N) \) exceeds\\ a predefined threshold \( \tau \), that remaining patent is marked as negative.\\ The process repeats iteratively. Newly marked negative patents are moved to \( M^N \), and the model \\re-evaluates the remaining patents in \( M^R \). The iteration stops when the number of remaining\\ patents \( M^R \) is less than a threshold \( \theta \), or no new patents are marked as negative.\\ The final output is a small set of patents in \( M^R \) that are \\not marked as negative, indicating their higher potential for technological innovation based \\on distinctiveness and low similarity to the negative patents.}
&  Various\\ \hline
\end{tabular}
}
\caption{This Table lists the various novelty indices brought by the literature. As noted in \citet{lanjouw2004patent}, however, no single indicator can fully capture the quality or value of a patent and multiple indicators (citations, claims, oppositions, family size) provide a more comprehensive and nuanced understanding of both the technological importance and the economic potential of innovations.}
\label{tab:innovation indices}
\end{table}
Given the richness of the literature about the subject, as aforementioned, the papers that address ``patent novelty'' adopt their own definitions of the term. Consequently, if we attempt to classify the main streams of literature on patent novelty, we may identify the following:
\begin{itemize}
\item[i)] The older and more established literature is the one that defines novelty basing on the forward and backward citations \citep{trajtenberg1997university, squicciarini2013measuring}. 
Moreover, in the document about PATSTAT database we can read: ``The number of citing patent documents can be an indicator of the importance of the patent. A frequently cited patent can be an indication of a core technology''.  
This literature have been recently criticized. In \citet{abbas2014literature}, for instance, the authors primarily discuss patent novelty in the context of citation analysis, classification systems, and keywords. They highlight how backward citations (patents cited by a new patent) are commonly used to infer novelty—specifically, fewer backward citations can signal higher novelty because the patent draws less from existing technologies. However, they criticize citation-based methods, noting that they don't always capture the full semantic or functional novelty of a patent.

\item[ii)] In \citet{fleming2001recombinant} novelty is primarily defined through the recombination of existing knowledge. The core idea is that innovations arise by combining different components (Fleming uses U.S. patent classification codes as a proxy for these components by examining how frequently different combinations of these classifications appear together in patents, Fleming can
assess the novelty of the recombination\footnote{Specifically, he looks at how often certain technological combinations have been used before. If a patent combines classifications that rarely appear together or haven't been combined before, it's considered more novel. This method helps to quantify novelty based on how ``new'' or ``unusual'' the combination of technological elements is, rather than just looking at the number of citations or backward links. 
Fleming also explores the uncertainty associated with novelty, showing that the more novel the combination (i.e., the less frequently those components have been combined in the past), the greater the uncertainty in terms of the patent's success.}) in new ways. A patent is considered novel when it brings together elements (or knowledge) that haven't been combined before, thus creating something new. Fleming emphasizes that recombinant novelty can vary in terms of uncertainty and outcomes: some combinations lead to breakthroughs, while others fail to generate impactful results.  
Fleming also makes a distinction between simple recombination, which involves putting together elements that are closely related or have been combined before, and more radical recombination, where the components are from different, often unrelated, technological fields. The latter leads to higher novelty and a greater chance of disruption but also comes with more uncertainty in terms of success. Thus, novelty in this context is seen as the creation of new combinations of existing components, with the degree of novelty depending on how distant or different these components are from one another. 

\item[iii)] In the early $2000$s also another stream of literature developed following the influential publication of \citet{aghion2005competition} according to which novelties were characterized by a certain degree of creative destruction. This idea has been incorporated in some studies on patents the most insightful of which is maybe \citet{autor2020foreign}. The idea, though not explicitly stated, is that a novel patent is one that ``destroys'' similar patents (e.g. reducing the sales of products sold with patents in the same sector). Though this idea is very appealing it cannot be easily implemented.

\item[iv)] The most recent stream is the one of Natural Language Programming (NPL) and text analyses of patents \citep{abbas2014literature,gerken2012new}. In essence in \citet{gerken2012new} the process of calculating patent novelty compares the text of a new patent with all previously filed patents. The similarity between the new patent and each prior one is measured, and the highest similarity score (the patent's ``oldness'') is subtracted from $1$ to determine the novelty score. Essentially, the more distinct a new patent is from earlier patents, the higher its novelty. This approach emphasizes the unique content of a patent by minimizing similarities with existing technologies. 
    In \citet{abbas2014literature}, semantic analysis is mentioned as a developing method, with an emphasis on how linguistic analysis (e.g., analyzing keywords, abstracts) is becoming crucial in identifying novel patents. But, their definition of novelty primarily revolves around citation gaps\footnote{If a patent doesn't reference many previous patents (or none at all), it suggests that the new patent is introducing ideas not heavily reliant on earlier work. This is seen as a sign of novelty because it indicates that the patent isn't just building on existing knowledge, but possibly presenting something new.} and technological distance\footnote{This refers to how different a patent is from those in similar fields. If a patent belongs to a new or less crowded technological class (based on how patents are categorized), it suggests that the patent is more novel because it explores a technological area that hasn't been fully developed yet.} (e.g., patents in new technological classes). 
    More recently \citet{kelly2021measuring} defined, again, patent novelty by focusing on textual similarity. They measure novelty by comparing the text content of patents to both previous and future patents. A patent is considered novel if it is distinct from earlier patents but also relevant to future innovations. Their approach uses textual analysis to identify patents that contribute significantly to technological progress, and these are often predictive of future citations and market value.
\end{itemize}
Table \ref{tab:innovation indices} summarizes the above indices. Note that different indices may capture different aspects of a patent. For example, the forward citations of a patent $n$ measure its impact on subsequent ``production'' or innovation, whereas the composition of patent $n$ across categories reflects its degree of ``interdisciplinarity'', which can be interpreted as a form of ``originality''. Therefore, it is not necessarily the case that all indices are positively correlated with one another.

In the main text we adopt a category-version of the \citet{squicciarini2013measuring} index as our primary measure of patent novelty due to its widespread recognition and methodological robustness within the economics of innovation literature \citep{higham2021patent, dehghani2023assessing}. This index, endorsed and implemented by the OECD in its patent quality indicators framework, captures the diversity of technological fields cited by a patent-interpreted as a proxy for knowledge recombination and inventive originality. It has also been applied across a range of recent empirical studies \citep{dehghani2023assessing, gao2023ex, angori2024patent} reflecting its growing standardization and comparability across datasets and time. Compared to newer text-based or machine-learning approaches, the \citet{squicciarini2013measuring}'s index offers a better replicability, and ease of modification in a cross-category setting as ours. Notice, however that our setting is general enough to allow for other measures of patent novelty as well.

%%%%%%%%%%%%%%%%%%%%%%%%%%%%%%%%%%%%%%%%%%%%%%%%%%%%%%%%%%%%%%%%%%%%%%%%%%%

\section{Proofs of the main results and technical details for the statistical test}
\label{app-proofs}

We here provide the rigorous proofs of the main results and the technical details that explain the performed statistical test. These proofs are based on the general theory we develop 
in Subsection~\ref{app-general} and which has the merit to provide a 
general mathematical framework that can be applied also in other settings. 

\begin{proof}[Proof of Theorem \ref{th-P}]
For each time step $t$, set ${\vX}_t=(X_{t,1},\dots,X_{t,N})^\top$, ${\vP}_t=(P_{t,1},\dots,P_{t,N})^\top$ and ${\vtheta}=(\theta_{1},\dots,\theta_{N})^\top$. Hence, 
the vectorial dynamics for the random vector ${\vP}_t$ is, for $t\geq 0$,
\begin{equation}\label{eq-dynamics-vector}
  \begin{aligned}
    {\vP}_0&=\vtheta/c \neq \vzero\quad\text{a.s.}\\
    {\vP}_{t+1}&=
    \Big(1-\frac{1}{t+1}\Big){\vP}_{t}+\frac{1}{t+1} \Gamma^\top {\vP}_{t+1} + O(1/t^2)\vone
   \\ &=
    {\vP}_t - \frac{1}{t+1}(I-\Gamma^\top){\vP}_t + 
    \frac{1}{t+1}\Gamma^\top \Delta {\vM}'_{t+1} +
    O(1/t^2)\vone ,
\end{aligned}
\end{equation}
where $\Delta{\vM}'_{t+1}={\vX}_{t+1}-{\vP}_t$.  
\\
Now, we fix $x>0$ and set
\begin{equation}\label{eq:zeta}
\zeta_0(x)=1,\qquad \zeta_{t}(x)=
\frac{\Gamma(t+x)}{\Gamma(t)}\sim
t^{x}\uparrow +\infty \,.
\end{equation}
More precisely, from  \cite[Lemma 4.1]{Gouet93} we have
$$
\zeta_t(x)=t^x + O(t^{x-1})\quad\mbox{and}\quad \frac{1}{\zeta_t(x)}=\frac{1}{t^x}+O(1/t^{x+1}),
$$
and so 
\begin{equation}\label{eq-ordine}
\begin{aligned}
&\frac{1}{\zeta_{t+1}(x)\zeta_{t+1}(1-x)}= \\
&\Big( \frac{1}{(t+1)^x}+O(1/t^{x+1}) \Big)
\Big( \frac{1}{(t+1)^{1-x}}+O(1/t^{2-x}) \Big)
=\\
&
\frac{1}{t+1}+O(1/t^2).
\end{aligned}
\end{equation}

\indent Hence,  multiplying \eqref{eq-dynamics-vector} by $\zeta_{t+1}(1-\gamma^*)$ and using the relation 
$$
\frac{\zeta_{t+1}(x)}{\zeta_t(x)}=
\frac{\Gamma(t+x+1)}{\Gamma(t+1)}
\frac{\Gamma(t)}{\Gamma(t+x)}=
1+\frac{x}{t}=
1+\frac{x}{t+1}+O(1/t^2),$$
with $x=1-\gamma^*$, 
 we get the following dynamics for ${\vA}_{t}=\zeta_{t}(1-\gamma^*){\vP}_{t}$, where we set 
 $\Delta{\vM}_{t+1}=\zeta_{t+1}(1-\gamma^*)\Delta{\vM}'_{t+1}$: 
$$
\begin{aligned}
{\vA}_{t+1}&=
\Big(1-\frac{1}{t+1}(I-\Gamma^\top)\Big)\frac{\zeta_{t+1}(1-\gamma^*)}{\zeta_t(1-\gamma^*)}{\vA}_{t}
+ \frac{1}{t+1}\Gamma^\top\Delta {\vM}_{t+1}
+O\Big(\frac{\zeta_{t+1}(1-\gamma^*)}{t^2}\Big)\vone
\\
&=
\Big(1-\frac{1}{t+1}(I-\Gamma^\top)\Big)
\Big(1+\frac{1-\gamma^*}{t+1}+O(1/t^2)\Big){\vA}_{t}
+ \frac{1}{t+1}\Gamma^\top\Delta {\vM}_{t+1}
+O\Big(\frac{\zeta_{t+1}(1-\gamma^*)}{t^2}\Big)\vone
\\
&=
{\vA}_{t}-\frac{1}{t+1}(\gamma^* I-\Gamma^\top){\vA}_{t}
+ \frac{1}{t+1}\Gamma^\top\Delta {\vM}_{t+1}+O\Big(\frac{\zeta_{t+1}(1-\gamma^*)}{t^2}\Big)\vone\,,
\end{aligned}
$$
with $ O(\zeta_t(1-\gamma^*)/t^2)=O(1/t^{1+\gamma^*})$. 
\\

Now, we are going to apply Theorem \ref{th-general-as-2} of Section~\ref{app-general} with $\Phi=\Gamma$, $\phi^*=\gamma^*$ and 
${\mathcal F}=({\mathcal F}_t)_t$ the natural filtration associated to the model, i.e.~${\mathcal F}_t=\sigma(X_{n,h}:\, n\leq t,\, h=1,\dots, N)$. To this purpose, we choose $\vu$ and $\vv$ as the 
left and the right eigenvectors of $\Gamma$ associated to $\gamma^*$, with strictly positive components and
such that $\vv^\top \vone = 1$ and $\vv^\top \vu = 1$. (Recall that $\gamma^*$ is simple and it is possible to choose the  components of these
vectors all strictly positive because of the Frobenius-Perron theory.) We set 
$\widetilde{A}_t=\vv^\top\vA_t=\zeta_t(1-\gamma^*)\vv^\top\vP_t$.
 First of all, we observe that assumptions (i) and (ii) of Theorem \ref{th-general-as-2} are satisfied because of {\Asa} and {\Asb}, that imply $\gamma^*\in (0,1]$. Moreover, also assumption (iii) is verified: indeed, we have 
 $\sup_{t}E[\widetilde{A}_t]<+\infty$ as $\widetilde{A}_t$ is non-negative (see Remark \ref{rem:non-negative}).  Finally, we have 
$$W_t=
\sum_{j=1}^N E[(\Delta M_{t+1,j})^{2}\,|\,\mathcal{F}_t]=
\zeta_{t+1}(1-\gamma^*)^2\sum_{j=1}^N   P_{j,t}(1-P_{j,t})\leq 
\zeta_{t+1}(1-\gamma^*)^2 \sum_{j=1}^N  P_{j,t}.$$
Then, denoting by $v_{\min}>0$ the minimum element of $\vv$, we obtain 
$$
\zeta_{t+1}(1-\gamma^*)^2\sum_{j=1}^N   P_{j,t}\leq 
\zeta_{t+1}(1-\gamma^*)^2 \vv^\top \vP_t/v_{\min} \leq 
2\zeta_{t+1}(1-\gamma^*)\widetilde{A}_t/v_{\min}.
$$
and so 
$$
W_t\leq \frac{1}{\gamma^*v_{\min}} 
\zeta_{t+1}(1-\gamma^*) \widetilde{A}_t\,.
$$ 
Therefore, recalling that $\sup_{t}E[\widetilde{A}_t]<+\infty$ as told before, 
we get  
$$
\sum_t \frac{1}{(t+1)^2} E[W_t] \leq 
\frac{1}{\gamma^*v_{\min}}\sup_{t}E[\widetilde{A}_t] \sum_t \frac{\zeta_{t+1}(1-\gamma^*)}{(t+1)^2} < +\infty.
$$
Hence, also assumption (iv) in Theorem \ref{th-general-as-2} is satisfied and we get 
$$
t^{1-\gamma^*} \vP_t\stackrel{a.s./L^2}\sim \zeta_t(1-\gamma^*)\vP_t=\vA_t \stackrel{a.s.}\longrightarrow \vP_{\infty}=\widetilde{P}_\infty \vu\,,
$$
where $\widetilde{P}_\infty=\widetilde{A}_\infty$ is a square-integrable non-negative random variable, which is the almost sure limit of the above defined  
$\widetilde{A}_t$, or equivalently (by \eqref{eq:zeta}), of $t^{1-\gamma^*}\vv^\top \vP_t$ . 
\\

It remains to prove that $P(\widetilde{P}_\infty>0)=1$. To this purpose, we use \cite[Theorem~S1.3]{ale-cri-ghi-innovation-2023}. Indeed, setting $\widetilde{P}_t=\vv^\top \vP_t$, we have 
\begin{equation}\label{widetildeP-dyn}
\begin{aligned}
\widetilde{P}_0 &=\vv^\top \vP_0 = \frac{1}{c} \vv^\top \boldsymbol{\theta}>0\\
\widetilde{P}_{t+1} &=\Big(1-\frac{1}{t+1}\Big)\widetilde{P}_t+
\frac{\gamma^*}{t+1}\widetilde{X}_{t+1} + O(1/t^2),\quad t\geq 0,
\end{aligned}
\end{equation}
with $\widetilde{X}_{t+1}=\vv^\top \vX_{t+1}$. Hence, 
if we define the stochastic process $\mathcal{V}=(\mathcal{V}_t)_{t\geq 0}$, 
taking values in the interval $[0,1]$, as 
$$
\begin{aligned}
\mathcal{V}_0 &=\widetilde{P}_0>0,
% \\
\qquad \qquad 
\mathcal{V}_{t+1} 
% &
=
\Big(1-\frac{1}{t+2}\Big)\mathcal{V}_t+\frac{1}{t+2} Y_{t+1},\quad t\geq 0,
\end{aligned}
$$
where $Y_{t+1}=\gamma^*\widetilde{X}_{t+1}$ (that takes values in $[0,1]$, since $0<\gamma^*\leq 1$,  
$X_{t+1,j}\in\{0,1\}$ and $\mathbf{v}^\top\mathbf{1}=1$), then by the technical result in  Subsection~\ref{technical-result-app} (applied to $\mathcal{W}_t=\widetilde{P}_t$ with $\beta=1$) we have
$$
|\widetilde{P}_{t}-\mathcal{V}_{t}|=O(\ln(t)/t)\to 0
$$
and also
$$
t^{1-\gamma^*}|\widetilde{P}_{t}-\mathcal{V}_{t}|=O(t^{1-\gamma^*}\ln(t)/t)
=O(\ln(t)/t^{\gamma^*})\to 0\,.
$$
Hence, from \cite[Theorem~S1.3]{ale-cri-ghi-innovation-2023} applied to ${\mathcal V}=(\mathcal{V}_t)$ 
with $\delta=\gamma^*$ (note that $E[Y_{t+1}|\mathcal{F}_t]=\gamma^*\widetilde{P}_t=\gamma^*\mathcal{V}_t+O(\ln(t)/t)$, we get that $t^{1-\gamma^*}
\mathcal{V}_t$ converges 
almost surely to a strictly positive finite random variable. 
This random variable is obviously also the almost sure limit of 
$t^{1-\gamma^*}\widetilde{P}_t$ and so we can conclude that $P(\widetilde{P}_\infty>0)=1$.  
%\qed
\end{proof}

\begin{proof}[Proof of Theorem~\ref{th-S}]
From Theorem~\ref{th-P}, using the same notation adopted in its proof, we get  
$$
S_{t,h}=\sum_{n=1}^t X_{n,h}\qquad\mbox{with } 
E[X_{t+1,h}\,|\,{\mathcal F}_t]=P_{t,h}\stackrel{a.s.}\sim \frac{ P_{\infty,h} }{t^{1-\gamma^*}}
$$
and so, by \cite[sec.~12.15]{williams},  we get 
\begin{equation}\label{eq-conv-as-S}
S_{t,h}\stackrel{a.s.}\sim S_{\infty,h}\,t^{\gamma^*}~\mbox{ with }
S_{\infty,h}=\frac{ P_{\infty,h} }{\gamma^*}\,.
\end{equation}
(We can also note that, similarly,  since the convergence in Theorem~\ref{th-P} is also in mean, we also have $E[S_{t,h}]=\sum_{n=0}^t E[X_{n,h}]=\sum_{n=0}^t E[P_{n,h}]\sim t^{\gamma^*}E[S_{\infty,h}]$.)  As a consequence of \eqref{eq-conv-as-S} and the fact that $P_{\infty,h}/P_{\infty,j}=u_h/u_j$, 
we obtain 
$$
\frac{ S_{t,h} }{ S_{t,j} }\stackrel{a.s.}\longrightarrow \frac{ S_{\infty,h} }{ S_{\infty,j} }=\frac{u_h}{u_j}
\,.
$$
\end{proof}

\begin{proof}[Proof of Theorem~\ref{cor-X}] 
In the proof of Theorem~\ref{th-P}, we have proven that the limit random vector of $t^{1-\gamma^*}\vP_{t}$ is $\vP_{\infty}=\widetilde{P}_\infty \vu$, where here $\widetilde{P}_\infty$ refers to a precise choice for the vector $\vu$, that is $\widetilde{P}_\infty=\widetilde{P}_\infty(\vu)$.  If we choose a different (left) eigenvector $\vu'$ of $\Gamma$ associated to $\gamma^*$, then we necessarily have have $\vu'=C\vu$ with $C\neq 0$ and so we can write the limit random vector as $\vP_{\infty}=\widetilde{P}_\infty(\vu')\vu'$ with 
$\widetilde{P}_\infty(\vu')=\widetilde{P}_\infty(\vu)/C$. Summing up, for any choice of the (left) eigenvector $\vu$, the limit random vector can be written as $\vP_{\infty}=\widetilde{P}_\infty(\vu)\vu$ with a suitable square-integrable random variable $\widetilde{P}_\infty(\vu)$ such that 
$P(\widetilde{P}_\infty(\vu)\neq 0)=1$ and 
$\widetilde{P}_\infty(C\vu)=\widetilde{P}_\infty(\vu)/C$. In other words, each random variable $P_{\infty,h}$ can be factorized in the product of a deterministic term  specific for each $h$, i.e.~$u_h$, and a common random term, i.e. $\widetilde{P}_\infty(\vu)$. 
\\
\indent Since $X_{t+1,h}\in\{0,1\}$ with $E[X_{t+1,h}\,|\,{\mathcal F}_t]=P_{t,h}$ and the convergence of 
$t^{1-\gamma^*}\vP_{t}$ toward $\vP_{\infty}$ is also in quadratic mean, we obtain 
\begin{multline*}
t^{1-\gamma^*} Var[X_{t+1,h}]= t^{1-\gamma^*} E[P_{t,h}(1-P_{t,h})] 
\\
\longrightarrow
\begin{cases}
E[P_{\infty,h}]=u_h E[\widetilde{P}_\infty(\vu)] \; &\text{for } \gamma^*<1 \\
E[P_{\infty,h}(1-P_{\infty,h})]=u_hE\left[\widetilde{P}_{\infty}(\vu)(1-u_h\widetilde{P}_\infty(\vu))\right]\; &\text{for } \gamma^*=1 \,.
%\\
%&(\text{with } \vu=\vone)
\end{cases}
\end{multline*}
Therefore, since the sign of the components of $\vu$ (which is the same for all of them) necessarily coincides with the one of $\widetilde{P}_\infty(\vu)$, in order to obtain the first limit relation in Corollary~\ref{cor-X}, we can set $\alpha(\vu)$ equal to $|E[\widetilde{P}_\infty(u)]|$ when $\gamma^*<1$ and equal to $|E\left[\widetilde{P}_{\infty}(\vu)(1-u_h\widetilde{P}_\infty(\vu))\right]|$ when $\gamma^*=1$, so that the two possible cases in the above formula can be summarized as $|u_h|\alpha(\vu)$.  Moreover, for each pair $h\neq j$, since  $X_{t+1,h}$ and $X_{t+1,j}$ are conditionally independent given ${\mathcal F}_t$, we get  $$
t^{2(1-\gamma^*)}cov(X_{t+1,h}, X_{t+1,j})=t^{2(1-\gamma^*)}cov(P_{t,h}, P_{t,j})\longrightarrow  
cov(P_{\infty, h}, P_{\infty, j})= u_h u_jVar[\widetilde{P}_{\infty}(\vu)]\,. 
$$
Hence, in order to obtain the second desired relation, we can set 
$\sigma^2(\vu)=Var[\widetilde{P}_{\infty}(\vu)]$. Finally, as a consequence, for the correlation coefficients, we have 
$$
t^{1-\gamma^*} \rho(X_{t+1,h},X_{t+1,j})=\frac{t^{2(1-\gamma^*)}cov(X_{t+1,h},X_{t+1,j})}{\sqrt{t^{(1-\gamma^*)}Var[X_{t+1,h}]}\sqrt{t^{(1-\gamma^*)}Var[X_{t+1,j}]}}\longrightarrow \sqrt{u_hu_j} \tfrac{\sigma^2(\vu)}{\alpha(\vu)}\,. 
$$
\end{proof}

\begin{proof}[Proof of Theorem~\ref{th-S-2}]
From Theorem~\ref{th-P} and Theorem~\ref{th-S}, using the same notation adopted in their proofs, we have by {\Asc}
$$
\begin{aligned}
E[X_{t+1,h}Y_{t+1,k}
% & 
\,|\, {\mathcal F}_t] 
% \\ 
&  =
E[Y_{t+1,k}] E[X_{t+1,h}\,|\, {\mathcal F}_t]=
\pi_k P_{t,h}\\
&\stackrel{a.s.}\sim t^{-(1-\gamma^*)} \pi_k P_{\infty,h}  
% \\ &
= 
t^{-(1-\gamma^*)}\pi_k \gamma^* S_{\infty,h}
\end{aligned}
$$
and so it is enough to apply \cite[sec.~12.15]{williams} in order to obtain 
$S_{t,k,h}=\sum_{n=1}^t X_{n,h}Y_{n,k}\stackrel{a.s.}\sim \pi_k S_{\infty,h} \, t^{\gamma^*}$. 
\end{proof}

\subsection*{Technical details for the statistical inference (Subsec.~\ref{stat-inference})}
We here use the same choice of the eigenvectors $\vv$ and $\vu$ and the same notation adopted in the proof of Theorem~\ref{th-P}.  
We recall that the dynamics of the above defined process $\vA_t = \zeta_t(1 - \gamma^*)\vP_t$ with $\zeta_t(\cdot)$ defined in \eqref{eq:zeta}, is 
$$ % \begin{equation*}
\begin{aligned}
\vA_0&=\vtheta/c
\\
{\vA}_{t+1}&=
{\vA}_{t}-\frac{1}{t+1}(\gamma^* I-\Gamma^\top){\vA}_{t}
% \\ & \qquad 
+
\frac{1}{t+1}\Gamma^\top\Delta {\vM}_{t+1}+O\left(\frac{\zeta_{t+1}(1-\gamma^*)}{t^2}\right)\vone\,,
\end{aligned}
$$ % \end{equation*}
where $ O(\zeta_t(1-\gamma^*)/t^2)=O(1/t^{1+\gamma^*})$ and $\Delta{\vM}_{t+1}=\zeta_{t+1}(1-\gamma^*)\Delta{\vM}'_{t+1}$ with $\Delta{\vM}'_{t+1}={\vX}_{t+1}-{\vP}_t$. Moreover, 
setting $\vS_t=(S_{t,1},\dots,S_{t,N})$ and $\vB_t=\frac{1}{\zeta_{t}(\gamma^*)}\vS_t$, we find the following vectorial dynamics: 
$$ % \begin{equation*}
\begin{aligned}
\vB_0&=\vzero\\
\vB_{t+1}&=
\frac{\zeta_t(\gamma^*)}{\zeta_{t+1}(\gamma^*)}\vB_t+
\frac{1}{\zeta_{t+1}(\gamma^*)}\vX_{t+1}
% \\ &
=
\Big(1-\frac{\gamma^*}{t+1}\Big) \vB_t+
\frac{1}{\zeta_{t+1}(\gamma^*)}\Delta \vM'_{t+1}+\frac{1}{\zeta_{t+1}(\gamma^*)}\vP_t
\\ &
=\Big( 1-\frac{\gamma^*}{t+1} \Big) \vB_t+
\frac{1}{\zeta_{t+1}(\gamma^*)\zeta_{t+1}(1-\gamma^*)} \Delta\vM_{t+1}
% \\ &
+ \frac{1}{\zeta_{t+1}(\gamma^*)\zeta_{t+1}(1-\gamma^*)}\frac{\zeta_{t+1}(1-\gamma^*)}{\zeta_t(1-\gamma^*)} \vA_t\,.
\end{aligned}
$$ % \end{equation*}
Using \eqref{eq-ordine} and the relation $\zeta_{t+1}(x)/\zeta_t(x)=1+O(1/t)$,  we obtain 
\begin{equation}\label{eq-system-1}
\begin{aligned}
\vB_{t+1}&=
\vB_t- \frac{1}{t+1}\left(\gamma^*\vB_t-\vA_t \right)
\\
& \qquad +
\frac{1}{t+1} \Delta\vM_{t+1} + O(\zeta_t(1-\gamma^*)/t^2)\vone\,,
\end{aligned}
\end{equation}
where again $ O(\zeta_t(1-\gamma^*)/t^2)=O(1/t^{1+\gamma^*})$.  Finally, we observe that, by  \eqref{eq:zeta}, we have 
\begin{multline*}
t^{-(1-\gamma^*)} 
% & 
E[\Delta\vM_{t+1}\Delta\vM_{t+1}^\top\,|\mathcal{F}_t]
% \\ 
\ \stackrel{a.s.}\sim \ t^{1-\gamma^*}diag\left(P_{t,1}(1-P_{1,t}),\dots, P_{t,N}(1-P_{t,N})\right)
\\
\begin{aligned}
&\stackrel{a.s.}\longrightarrow
\begin{cases}
diag(\vP_\infty)\quad &\mbox{for } \gamma^*<1\\
diag(P_{\infty,1}(1-P_{\infty,1}),\dots,P_{\infty,N}(1-P_{\infty,N})) \quad &\mbox{for } \gamma^*=1\;(\mbox{and so } \vu=\vone) 
\end{cases}
\\
&\quad =
\begin{cases}
\widetilde{P}_\infty diag(\vu)\quad &\mbox{for } \gamma^*<1\\
\widetilde{P}_\infty (1-\widetilde{P}_\infty) I \quad &\mbox{for } \gamma^*=1\;(\mbox{and so } \vu=\vone) 
\end{cases}
\end{aligned}
\end{multline*}
and we recall that, as proven before, 
$\widetilde{A}_t=\vv^{\top}\vA_t=\zeta_t(1-\gamma^*)\vv^\top\vP_t\stackrel{a.s.}\to \widetilde{P}_\infty$. 
Therefore, the pair 
$(\vA_t,\vB_t)_t$ satisfies the dynamics and the conditions required in the general central limit theorem proven in~\cite[Appendix~A]{ale-cri-ghi-innovation-stat}, provided we assume 
${\mathcal Re}(\gamma^*_2)/\gamma^*<1/2$, where $\gamma^*_2$ is an eigenvalue of $\Gamma$ different from $\gamma^*$ with highest real part, that is $\gamma^*_2\in Sp(\Gamma)\setminus \{\gamma^*\}$ with 
 ${\mathcal Re}(\gamma^*_2)=\max\{ {\mathcal Re}(\gamma) : 
 \gamma \in Sp(\Gamma)\setminus \{\gamma^*\} \}$. 
Hence, we can apply the statistical tools based on that result and described in~\cite[Appendix~B and~C]{ale-cri-ghi-innovation-stat}. In particular, in the mean-field case we have $\vv=N^{-1}\vone$, 
$\vu=\vone$ and $\gamma^*_2=\gamma^*(1-\iota)$ so that, when  $\iota>1/2$, we have 
$$
 (2\iota-1)\frac{\|\vS_t-\vv^{\top}\vS_t\|^2}{\vv^\top\vS_t}
=(2\iota-1)\zeta_t(\gamma^*)
\frac{\|\vB_t-\vv^{\top}\vB_t\|^2}{\vv^\top\vB_t}
\stackrel{a.s.}\sim 
(2\iota-1)t^{\gamma^*}\frac{\|\vB_t-\vv^{\top}\vB_t\|^2}{\vv^\top\vB_t}\stackrel{d}\longrightarrow \chi^2(N-1).
$$

%%%%%%%%%%%%%%%%%%%%%%%%%%%%%%%%%%%%%%%%%%%%%%%%%%%%%%%%%%%%%%%%%%%%%%%%%%%%

\subsection{General results}\label{app-general}
Let ${\vA}_t=(A_{t,1},\dots,A_{t,N})^\top$, with $t\geq 0$, be a multi-dimensional real stochastic processes, adapted to a filtration $(\mathcal{F}_t)_t$, with the following dynamics:
\begin{equation}\label{eq-general-dynamics}
\vA_{t+1}=\vA_t - \frac{1}{t+1}(\phi^{*} I-\Phi^\top)\vA_t + \frac{1}{t+1} \Phi^\top \Delta {\vM}_{t+1} + \vR_{A,t+1}\\
\end{equation}
where  $\vA_0$ is {\em integrable} and 
\begin{itemize}
\item[$(i)$] $\Phi^\top$ is a non-negative irreducible matrix with leading eigenvalue $0<\phi^{*}\leq 1$;
\item[$(ii)$] $\vR_{A,t+1}=O(t^{-(1+\beta)})\vone$
for some $\beta > 0$.\footnote{The notation $R_t=O(s_t)$ means that $R_t$ is a (possibly random) reminder term such that $|R_t|\leq C s_t$ for a suitable constant $C$ and for $t$ large enough.}
\end{itemize}
Let $\vu$ and $\vv$ be the left and the right eigenvectors of $\Phi$ associated to $\phi^*$, with strictly positive components and such that
$
\vv^\top\vone=1\quad \mbox{and}\quad \vv^\top\vu=1.
$ (Recall that  $\phi^*$ is real and simple and it is possible to choose the components of
these vectors all strictly positive because of the Frobenius-Perron theory). 
 Set $\widetilde{A}_t=\vv^\top \vA_t$. 

\begin{theorem} 
Under (i) and (ii) and assuming 
\begin{itemize}
\item[(iii)] $\sup_t E[\,|\widetilde{A}_t|\,]<+\infty$,
\end{itemize}
we have 
$
{\widetilde A}_t\stackrel{\text{a.s.}}\longrightarrow \widetilde{A}_\infty\,,
$
where $\widetilde{A}_\infty$ is an integrable random variable.   
 \\
Moreover, if we also assume 
\begin{itemize}
\item[$(iv)$] $\vA_0$ square-integrable and 
$\sum_t w_t/(t+1)^2<+\infty$ where 
$w_t=\sum_{h=1}^N E[( \Delta M_{t+1,j} )^2]$,
\end{itemize}
then we have $\sup_{t}E[(\widetilde{A}_t)^2]<+\infty$ and so 
$\widetilde{A}_t$ converges to $\widetilde{A}_\infty$ also in quadratic mean (i.e. in $L^2$). 
\end{theorem}
Note that (iii) is verified when $(\widetilde{A}_t)_t$ is uniformly integrable and in this case we 
also have that the convergence is in mean. 

\begin{proof}
By multiplying equation \eqref{eq-general-dynamics} by $\vv^\top$ we obtain
 \begin{equation}\label{eq-dynamics-A_tilde-II} 
\widetilde{A}_{t+1}=
\widetilde{A}_{t}+
\frac{1}{t+1}\phi^*\Delta \widetilde{M}_{t+1}+\vv^\top \vR_{A,t+1}\,,
\end{equation}
where $\Delta\widetilde{M}_{t+1}=\vv^\top\Delta\vM_{t+1}$. Setting 
$\breve{M}_{t}=\sum_{n=1}^t \frac{1}{n}\Delta\widetilde{M}_n$ and $\widetilde{R}_{A,t+1}=\vv^\top\vR_{A,t+1}=O(1/t^{1+\beta})$ (by (ii)), we have 
\begin{equation}\label{eq-breve-M}
\widetilde{A}_{t+1}-\widetilde{A}_0=\sum_{n=0}^t(\widetilde{A}_{n+1}-\widetilde{A}_n)=\phi^*\breve{M}_{t+1}+\sum_{n=0}^t \widetilde{R}_{A,n+1}\,.
\end{equation}
Hence, since $\sup_{t}E[\,|\widetilde{A}_t|\,]<+\infty$ (by (iii)) and $\sum_t 1/t^{1+\beta}<+\infty$,
we also have $\sup_t E[\,|\breve{M}_t|\,]<+\infty$.   Therefore, $(\breve{M}_t)_t$ is a martingale 
bounded in $L^1$ and so it converges almost surely to an integrable random variable $\breve{M}_\infty$.  It follows from \eqref{eq-breve-M} that 
$(\widetilde{A}_t)$ converges almost surely to an integrable random variable $\widetilde{A}_\infty$.
Moreover, we obtain 
\begin{equation}\label{eq-quadratic-mean-conv}
\begin{aligned}
(\widetilde{A}_t-\widetilde{A}_\infty)^2 & = 
\big(\phi^*(\breve{M}_t-\breve{M}_\infty) + \sum_{n\geq t} \widetilde{R}_{A,n+1}\big)^2\\
&=(\phi^*)^2(\breve{M}_t-\breve{M}_\infty)^2 + O(1/t^{\beta})\,.
\end{aligned}
\end{equation}
Now, we are going to prove that, under assumption (iv), we have $\sup_t(\widetilde{A}_t)^2<+\infty$ 
 so that (by \eqref{eq-breve-M}) $(\breve{M}_t)_t$ is a martingale bounded in $L^2$ and so $\breve{M}_\infty$ is square-integrable and $\breve{M}_t$ converges in quadratic mean to it. By \eqref{eq-quadratic-mean-conv} 
 this fact obviously implies that $\widetilde{A}_t$ converges in quadratic mean to $\widetilde{A}_\infty$. 
\\
\indent We observe that, from (iv) and the fact that 
$(\Delta \widetilde{M}_{t+1})^2\leq
C \sum_{j=1}^N (\Delta M_{t+1,j})^2$, we obtain 
$$
E\Big[\,\Big|\frac{\Delta\widetilde{M}_{t+1}}{t+1}\widetilde{R}_{A,t+1}\Big|\,\Big]^2
\leq E\Big[\,\frac{(\Delta\widetilde{M}_{t+1})^2}{(t+1)^2}\Big]E[(\widetilde{R}_{A,t+1})^2]
\leq C \frac{w_t}{(t+1)^2} O(1/t^{2(1+\beta)})
$$
and so 
$$
E\Big[\,\Big|\frac{\Delta\widetilde{M}_{t+1}}{t+1}\widetilde{R}_{A,t+1}\Big|\,\Big]=o(1/t^{1+\beta})\,.
$$
Therefore, from \eqref{eq-dynamics-A_tilde-II}, since 
 $\sup_t E[\widetilde{A}_t]<+\infty$, we get  
$$ % \begin{equation*}
\begin{aligned}
E[(\widetilde{A}_{t+1})^2]
&\leq
E[(\widetilde{A}_t)^2]+
(\phi^{*})^2C\frac{w_t}{(t+1)^2} 
% \\ &
+ 
O\Big(\frac{1}{t^{2(1+\beta)}}\Big) +
O\Big(\frac{\sup_tE[\widetilde{A}_t]}{t^{1+\beta}}\Big) + 
o\Big(\frac{1}{t^{1+\beta}}\Big) \,,
\\
&= E[(\widetilde{A}_t)^2]+
(\phi^{*})^2C\frac{w_t}{(t+1)^2}+
O(1/t^{1+\beta})\,.
\end{aligned}
$$ % \end{equation*}
Then, we find 
$$ % \begin{equation*}
\begin{aligned}
|E[(\widetilde{A}_t)^2]- E[(\widetilde{A}_0)^2]| & \leq 
\sum_{n=0}^{t-1} |E[(\widetilde{A}_{n+1})^2] - E[(\widetilde{A}_n)^2]|\\
&\leq
(\phi^{*})^2 C\sum_n \frac{w_n}{(n+1)^2}+ 
\sum_n O(1/n^{1+\beta})<+\infty\,,
\end{aligned}
$$ % \end{equation*}
where we have used (iv) in order to say that the first series is finite.
Therefore, assuming $\vA_0$ (and so $\widetilde{A}_0$) square-integrable, 
we have $\sup_{t}E[(\widetilde{A}_t)^2]<+\infty$.    
\end{proof}

\begin{remark}[non-negative case]\label{rem:non-negative}
Condition (iii) is verified when $\widetilde{A}_t$ is non-negative for each $t$. Indeed,  for each $t$, we have   
\begin{equation*}
\begin{aligned}
|E[\widetilde{A}_t]-E[\widetilde{A}_0]|
\leq 
\sum_{n=0}^{t-1} |E[\widetilde{A}_{n+1}] - E[\widetilde{A}_n]|\leq
\sum_n O(1/n^{1+\beta})
\end{aligned}
\end{equation*}
 and thus, since the last series is finite and $\widetilde{A}_0$ is integrable,  
 it follows $\sup_t E[\widetilde{A}_t]<+\infty$.  Hence, if 
$\widetilde{A}_t$ is non-negative, we can conclude that (iii) is verified. 
\end{remark}

\begin{theorem}\label{th-general-as-2}
Assuming (i), (ii), (iii) and (iv), we have 
$$
{\vA}_t\stackrel{\text{a.s.}/L^2}\longrightarrow \widetilde{A}_\infty{\vu}\,,
$$
where $\widetilde{A}_\infty$ is a square-integrable random variable. 
\end{theorem}
\begin{proof}
We firstly want to prove that we can neglect the term $R_{A,t+1}$ in the dynamics \eqref{eq-general-dynamics} of $\vA_t$. 
\\
\indent We recall that the matrix $\Phi^\top$ can be decomposed as 
$$\Phi^\top=\phi^{*}\vu\vv^\top+UDV^\top\,,$$  
where $D$ is the diagonal matrix whose elements
are the eigenvalues of $\Phi$ (i.e. of $\Phi^\top$) different from $\phi^*$ and 
 $U$ and $V$ denote the matrices whose columns are
the left (right) and the right (left) eigenvectors of $\Phi$ (of $\Phi^\top$, respectively) associated to these eigenvalues, so that we have 
\begin{equation}\label{matrices-prop}
  V^\top\vu=U^\top\vv=0,\quad V^\top U=U^\top V=I,\quad
  I=\vu\vv^\top + UV^\top\,.
\end{equation}
Therefore the  dynamics of $\vA_t$ can be rewritten as follows:
\begin{equation}\label{utile-1}
\begin{aligned}
\vA_{t+1} & =\Big(I - \frac{1}{t+1}U(I\phi^{*}-D)V^\top\Big)\vA_t 
\\ & \qquad 
+ 
\frac{1}{t+1} \Phi^\top \Delta {\vM}_{t+1} + {\vR}_{A,t+1}.
\end{aligned}
\end{equation}
Let $\alpha_j=\phi^{*}-\phi_j$ with $\phi_j$ eigenvalue of $\Phi$ different from $\phi^*$. 
Then ${\mathcal Re}(\alpha_j)>0$. Moreover, we have 
\begin{equation}\label{utile-2}
\begin{aligned}
\vA_{t+1} &=\ C_{m_0,t}\vA_{m_0} + 
\sum_{k=m_0}^t\frac{1}{k+1}C_{k+1,t}\Phi^\top \Delta\vM_{k+1} 
\\ & \qquad 
+
\sum_{k=m_0}^t C_{k+1,t} \vR_{\vA,k+1} 
\\
&=\ 
C_{m_0,t}\vA_{m_0} + 
\sum_{k=m_0}^t\frac{1}{k+1}C_{k+1,t}\Phi^\top \Delta\vM_{k+1} + \boldsymbol{{\rho}}_{t+1}\,,
\end{aligned}
\end{equation}
where $m_0$ is such that ${\mathcal Re}(\alpha_j)/(m_0+1)<1$ for each $j$ and 
$$
\begin{aligned}
C_{k+1,t} &=  
\prod_{m=k+1}^t\Big(I-\frac{1}{m+1}(I\phi^{*}-\Phi^\top)\Big)
% \\ & 
=\prod_{m=k+1}^t\Big(I-\frac{1}{m+1}U(I\phi^{*}-D)V^\top\Big)
\\ &
= 
U\bigg(\prod_{m=k+1}^t\Big(I-\frac{1}{m+1}(I\phi^{*}-D)\Big)\bigg)V^\top,
\end{aligned}$$
and so we can write $ C_{k+1,t} = U A_{k+1,t} V^{\top} $ and
\begin{equation*}
%C_{k+1,t} = U A_{k+1,t} V^{\top},
%\qquad\mbox{and}\qquad
A_{k+1,t} = 
\prod_{m=k+1}^t \Big(I-\frac{1}{m+1}(I\phi^{*}-D)\Big).
\end{equation*}
Moreover, setting for any $x\in{\mathbb C}$ with ${\mathcal Re}(x)/(m_0+1)<1$, $p_{m_0}(x)=1$ and $p_k(x)=\prod_{m=m_0}^{k}(1-\tfrac{x}{m+1})$ for $k\geq m_0$ and 
$F_{k+1,t}=\frac{p_t(x)}{p_k(x)}$ for $m_0-1\leq k\leq t-1$, 
from \cite[Lemma A.5]{ale-cri-ghi-MEAN} we get
$$[A_{k+1,t}]_{jj} = F_{k+1,t}(\alpha_j)\,.$$
\indent We now prove that
$|\boldsymbol{{\rho}}_{t+1}|\stackrel{a.s.}\to 0$. 
To this end, first notice that $O(|C_{k+1,t}|)=O(|A_{k+1,t}|)$ and, setting 
$a_2^*={\mathcal Re}(\alpha_2^*)=\phi^*-{\mathcal Re}(\phi^*_2)$ with $\phi_2^*$ eigenvalue of 
$\Phi$, different form $\phi^*$,  
such that ${\mathcal Re}(\phi_2^*)=\max_j\{{\mathcal Re}(\phi_j)\}$, 
we have (see \cite[Lemma A.4]{ale-cri-ghi})
$$
|A_{k+1,t}| =
O\Big(\frac{|p_t(\alpha_2^{*})|}{|p_k(\alpha_2^{*})|}\Big)
= O\Big(\Big(\frac{k}{t}\Big)^{a_2^{*}}\Big) 
\quad\mbox{for } k=m_0,\dots, t-1,
$$
and simply $|A_{t+1,t}|=O(1)$ for $k=t$.
Moreover, recalling that $\vR_{A,t+1}=O(t^{-(1+\beta)})\vone$ for some $\beta > 0$,  we have
$$
\begin{aligned}
|\boldsymbol{{\rho}}_{t+1}|&= 
\Big|
\sum_{k=m_0}^t  C_{k+1,t} \vR_{A,k+1}
\Big|
% \\ &
= 
O\Big(\sum_{k=m_0}^{t-1} \Big(\frac{k}{t}\Big)^{a_2^{*}} \frac{1}{k^{{1+\beta}}}\Big)
+O(1/t^{1+\beta})\\
&=
O\Big( \frac{1}{t^{a_2^{*}}}
\sum_{k=m_0}^{t-1} k^{a_2^{*}-1-\beta}\Big)+O(1/t^{1+\beta})\to 0,
\end{aligned}
$$
because $a_2^*>0$ and $\beta>0$.
\\
\indent Therefore, in all the sequel, without loss of generality,  we can assume that $\vA_t$ follows the dynamics \eqref{eq-general-dynamics} with $\vR_{A,t+1}=\vzero$. 

\smallskip

We now decompose the vectorial process ${\vA}_t$ by means of the Jordan representation of the matrix $\Phi$. 
Specifically, for any $\phi\in Sp(\Phi)\setminus \{\phi^*\}$, 
we can denote as $J_\phi$ the Jordan block and with $U_\phi$ and $V_\phi$ the matrices whose columns are, respectively, the
left and right (possibly generalized) eigenvectors of $\Phi$ associated to the eigenvalue $\phi$, i.e. 
$$\Phi V_\phi = V_\phi J_\phi
\qquad\text{and}\qquad
U_\phi^\top \Phi = J_\phi U_\phi^\top.$$
Then, we can consider the decomposition
$$
{\vA}_t=\widetilde{A}_t{\vu}+
\sum_{\phi\in Sp(\Phi)\setminus \{\phi^*\}}{\vA}_{\phi,t}\,,
$$
where $\widetilde{A}_t={\vv}^\top{\vA}_t$ (as defined above) and
${\vA}_{\phi,t}= U_\phi V_\phi^\top {\vA}_t$.
\\
\indent We have already proven the almost sure convergence and the convergence in $L^2$ for $\widetilde{A}_t$ under (i), (ii), (iii) and (iv).  
In the following steps we are going to show that, under (i), (ii) and (iv), 
 each ${\vA}_{\phi,t}$ 
converges almost surely and in $L^2$ to zero. In particular, this last task will be done
separately for the eigenvalues with $|\phi|<\phi^*$ and with 
$|\phi|=\phi^*$. Remember that the assumption that 
$\Phi$ (or, equivalently, $\Phi^\top$) is irreducible ensures that 
$\phi^*$ is real, simple and $|\phi|\leq \phi^*$
for any $\phi\in Sp(\Phi)$. Moreover, let us set 
$W_t=\sum_{h=1}^N E[( \Delta M_{t+1,j} )^2\,|\,{\mathcal F}_t]$ and observe that assumption (iv) 
means $E[\sum_t W_t/(t+1)^2]=\sum_t E[W_t]/(t+1)^2=\sum_t w_t/(1+t)^2<+\infty$, which also implies $\sum_t W_t/(t+1)^2<+\infty$ almost surely. \\
 
\subsubsection*{Study of ${\vA}_{\phi,t}$ with $|\phi|<\phi^*$.}
Let $\breve{\vA}_t=V_\phi^\top\vA_t$ and since 
${\vA}_{\phi,t}=U_\phi V_\phi^\top\vA_t=U_\phi \breve{\vA}_t$,
it is enough to prove that $\|\breve{\vA}_t\|^2$ converges a.s. and in $L^2$ to zero.
To this end, by multiplying equation \eqref{eq-general-dynamics} by $V_\phi^\top$, we have
$$
\breve{\vA}_{t+1} = 
\Big(I-\frac{1}{t+1}
(\phi^*I-J_\phi^\top)\Big)\breve{\vA}_t + 
\frac{1}{t+1}J_\phi^\top V_\phi^\top \Delta {\vM}_{t+1}.
$$
Then, since for any real matrix $Q$ we can write 
\begin{equation}\label{eq:cross_product_Delta_M}
\begin{aligned}
E[\Delta\vM_{t+1}^{\top}
Q
\Delta\vM_{t+1}|\mathcal{F}_t] 
& = 
\sum_{j=1}^N q_{jj}^2
E[(\Delta M_{j,t+1})^2|\mathcal{F}_t] 
\\
& \leq \max_j{q_{jj}^2} W_t,
\end{aligned}
\end{equation}
we have that
$$
\begin{aligned}
E[\|\breve{\vA}_{t+1}\|^2 & |\mathcal{F}_t] \\
&\leq 
\big\|\bigg( \Big(
1 -\frac{\phi^*}{t+1}\Big)I
+ \frac{1}{t+1}J_{\phi}
\bigg)\breve{\vA}_{t}\big\|^2 + 
\frac{1}{(t+1)^2}
\sum_{j=1}^N [\bar{V}_{\phi} 
\bar{J}_{\phi}
J_{\phi}^\top
V_{\phi}^\top]_{jj}^2 E[(\Delta M_{j,t+1})^2|\mathcal{F}_t]
\\
&\leq
\Big( 1 -
 \frac{\phi^*}{t+1}
+ \frac{\|J_{\phi}\|_{2,2}}{t+1}
\Big)^2\|\breve{\vA}_{t}\|^2 + 
\frac{1}{(t+1)^2}
\max_j\{[\bar{V}_{\phi} 
\bar{J}_{\phi}
J_{\phi}^\top
V_{\phi}^\top]^2_{jj}\}
W_t.
\end{aligned}
$$
Then, regarding the first term, we note that
$$
\Big( 1 -
 \frac{\phi^*}{t+1}
+ \frac{\|J_{\phi}\|_{2,2}}{t+1}
\Big)^2 \leq 
\Big( 1 -
 \frac{\phi^*}{t+1}
+ \frac{|\phi|+\phi^*}{2(t+1)}
\Big)^2 =
\Big( 1 -
 \frac{\phi^* - |\phi|}{2(t+1)}
\Big)^2,
$$
and so
$$
E[\|\breve{\vA}_{t+1}\|^2|\mathcal{F}_t] \leq
\Big( 1 -
 \frac{\phi^* - |\phi|}{2(t+1)}
\Big)^2 \|\breve{\vA}_{t}\|^2 + 
\frac{C}{(t+1)^2}W_t.
$$
Therefore, since $\phi^*> |\phi|$ and by (iv), 
the process $\|\breve{\vA}_{t}\|^2$ is a non-negative almost supermartingale 
so that it converges almost surely (see \cite{rob}).
Moreover, by applying the expectation, we obtain
$$
E[\|\breve{\vA}_{t+1}\|^2] \leq
\Big( 1 -
 \frac{\phi^* - |\phi|}{2(t+1)}
\Big)^2E[\|\breve{\vA}_{t}\|^2] + 
\frac{C}{(t+1)^2}E[W_t],
$$
which, since $\sum_t (\phi^* - |\phi|)/(t+1)=+\infty$, by (iv) and 
 \cite[Lemma S1.6]{ale-cri-ghi-innovation-2023}, 
 we can conclude that 
$\|\breve{\vA}_{t}\|\stackrel{a.s./ L^2}\longrightarrow 0$,
and hence $\breve{\vA}_{t}\stackrel{a.s. /L^2}\longrightarrow \vzero$.\\

\subsubsection*{Study of ${\vA}_{\phi,t}$ with $|\phi|=\phi^*$.} 
From the Frobenius-Perron theory, we know that each eigenvalue with maximum modulus is simple.
 Let us denote the corresponding right and left eigenvectors by $\vv_\phi$ and $\vu_\phi$ of $\Phi$. 
Then, set $a_{\phi,t}=\vv_\phi^\top\vA_t$ so that, since we have  
${\vA}_{\phi,t}=\vu_\phi \vv_\phi^\top\vA_t=\vu_\phi a_{\phi,t}$,
it is enough to prove that $|a_{\phi,t}|$ converges to zero almost surely and in $L^2$.
To this end, by multiplying equation \eqref{eq-general-dynamics} by $\vv_\phi^\top$, we have
$$
{a}_{\phi,t+1} = 
\Big(1-\frac{1}{t+1}
(\phi^*-\phi)\Big){a}_{\phi,t }+ \frac{\phi}{t+1}\vv_\phi^\top \Delta {\vM}_{t+1}.
$$
Then, using \eqref{eq:cross_product_Delta_M},
we have that
$$
\begin{aligned}
% &
E[|a_{\phi,t+1}|^2|\mathcal{F}_t] 
% \\
&\leq 
\Big|
1 -\frac{\phi^*}{t+1}
+ \frac{\phi}{t+1}
\Big|^2 |a_{\phi,t}|^2 + 
\frac{|\phi|^2}{(t+1)^2}
\sum_{j=1}^N|v_{j}|^2E[(\Delta M_{j,t+1})^2|\mathcal{F}_t]
\\
&\leq
\Big|
1 -\frac{\phi^*}{t+1}
+ \frac{\phi}{t+1}
\Big|^2 |a_{\phi,t}|^2  + 
\frac{|\phi|^2}{(t+1)^2}
\max_j\{|v_j|^2\}W_t.
\end{aligned}
$$
Then, regarding the first term we have that
$$
\begin{aligned}
\Big|
1 -\frac{\phi^*}{t+1}
+ \frac{\phi}{t+1}
\Big|^2
&=\Big(
1 -\frac{\phi^*}{t+1}
+ \frac{\mathcal{R}e(\phi)}{t+1}
\Big)^2 +
\Big( \frac{\mathcal{I}m(\phi)}{t+1}\Big)^2 \\
&=
1 +
 \Big( \frac{\phi^* - \mathcal{R}e(\phi)}{t+1}\Big)^2
  - 2 \Big(\frac{\phi^* - \mathcal{R}e(\phi)}{t+1}\Big)
 + \Big( \frac{\mathcal{I}m(\phi)}{t+1}\Big)^2 
\\
&=
1 - \Big(\frac{2(\phi^* - \mathcal{R}e(\phi))}{t+1}\Big)
  +  \Big( \frac{\phi^{*2} 
  -2\phi^{*}\mathcal{R}e(\phi) + \mathcal{R}e(\phi)^2+
  \mathcal{I}m(\phi)^2}{(t+1)^2}\Big)
\\
&=1 - \Big(\frac{2(\phi^* - \mathcal{R}e(\phi))}{t+1}\Big)
  +  \Big( \frac{2\phi^{*} (\phi^{*} -\mathcal{R}e(\phi))}{(t+1)^2}\Big)
\\
&=1 - 2\Big(\frac{1}{t+1} -
  \frac{\phi^{*}}{(t+1)^2}\Big)
  (\phi^* - \mathcal{R}e(\phi))
\\
\end{aligned}
$$
and so
$$
\begin{aligned}
E[|a_{\phi,t+1}|^2|\mathcal{F}_t] &\leq
\Big( 1 - 2\Big(\frac{1}{t+1} -
  \frac{\phi^{*}}{(t+1)^2}\Big)
  (\phi^* - \mathcal{R}e(\phi))
\Big) |a_{\phi,t}|^2 + 
\frac{C}{(t+1)^2}W_t.
\end{aligned}
$$
Therefore, since $\phi^*> \mathcal{R}e(\phi)$ and by (iv), 
the process $|a_{\phi,t}|^2$ is a non-negative almost supermartingale so that it 
converges almost surely (see \cite{rob}).
Moreover, by applying the expectation, we obtain
$$
\begin{aligned}
E\big[|a_{\phi,t+1}|^2\big] &\leq
\Big(1 - 2\Big(\frac{1}{t+1} -
  \frac{\phi^{*}}{(t+1)^2}\Big)
  (\phi^* - \mathcal{R}e(\phi))
\Big)E\big[|a_{\phi,t}|^2\big] + 
\frac{C}{(t+1)^2}E[W_t]\,.
\end{aligned}
$$
Since $\sum_t (1/(t+1)-\phi^*/(t+1)^2)=+\infty$ and by (iv) and 
\cite[Lemma S1.6]{ale-cri-ghi-innovation-2023},  
we can conclude that 
$|a_{\phi,t}|\stackrel{a.s./L^2}\longrightarrow 0$,
and hence $a_{\phi,t}\stackrel{a.s./L^2}\longrightarrow 0$.
\end{proof}
%%%%%%%%%%%%%%%%%%%%%%%%%%%%%%%%%%%%%%%%%%%%%%%%%%%%%%%%%%%%%

\subsection{Technical result}\label{technical-result-app}
Let ${\mathcal W}=(\mathcal{W}_t)_{t\geq 0}$ be a bounded stochastic process with the following dynamics 
$$ % \begin{equation}\label{rw-dyn-O}
\mathcal{W}_{t+1}=
\Big(1-\frac{1}{t+1}\Big)\mathcal{W}_t+\frac{1}{t+1} Y_{t+1}+R_{t+1},\quad t\geq 0,
$$ % \end{equation}
where $(Y_t)_t$ is a bounded stochastic process and $R_t=O(1/t^{1+\beta})$ with $\beta>0$. 
Define a bounded stochastic process $\mathcal{V}=(\mathcal{V}_t)_{t\geq 0}$  with dynamics   
$$ % \begin{equation}\label{rw-dyn-O}
\begin{aligned}
\mathcal{V}_{t+1}&=
\Big(1-\frac{1}{t+2}\Big)\mathcal{V}_t+\frac{1}{t+2} Y_{t+1},\quad t\geq 0\,.
\end{aligned}
$$ % \end{equation}
Then 
$$
|\mathcal{W}_t-\mathcal{V}_t|=
O(1/t)|\mathcal{W}_0-\mathcal{V}_0|+
%O\Big(\frac{\ln(t)}{t}\Big)+
\frac{1}{t}\sum_{n=0}^t O(n^{-\beta'})\,,
%\begin{cases}
%O(1/t) \quad\text{if } \beta>1\\
%O(\ln(t)/t) \quad\text{if } \beta=1\\
%O(1/t^\beta) \quad\text{if } \beta\in (0,1)\,.
%\end{cases}
$$
where $\beta'=\min(\beta,1)$.
% (We have to distinguish three cases: $\beta>1,\, =1,\, <1$, but in all the three cases, we can say that there exists $\beta'\in (0,1)$ such that $|\mathcal{W}_t-\mathcal{W}_t'|=O(1/t^{\beta'})\to 0$.)
% \\

\smallskip

Indeed, we can observe that  we can write 
\begin{equation*}
\mathcal{W}_{t+1}=
\Big(1-\frac{1}{t+2}\Big)\mathcal{W}_t+\frac{1}{t+2} Y_{t+1}+R'_{t+1},\quad t\geq 0,
\end{equation*}
with $R'_{t+1}=R_{t+1}+O(1/t^2)=O(1/t^{1+\beta'})$ and so 
$$
\mathcal{W}_{t+1}=C_{0,t}\mathcal{W}_0 + \sum_{n=0}^t C_{n+1,t}\frac{Y_{n+1}}{n+2} +
 \sum_{n=0}^t C_{n+1,t} R'_{n+1}
$$
where 
\begin{equation*}
\begin{aligned}
C_{0,t}&=\prod_{m=0}^t \Big(1-\frac{1}{m+2}\Big)=O(1/t)
\\
C_{n+1,t}&=\prod_{m=n+1}^t \Big(1-\frac{1}{m+2}\Big)=
\frac{\prod_{m=0}^t \big(1-\frac{1}{m+2}\big)}
{\prod_{m=0}^n \big(1-\frac{1}{m+2}\big)}=O(n/t)
\end{aligned}
\end{equation*}
Similarly, we have 
$$
\mathcal{V}_{t+1}=C_{0,t}\mathcal{V}_0 + \sum_{n=0}^t C_{n+1,t}\frac{Y_{n+1}}{n+2}.
$$
Hence, we obtain 
$$
\begin{aligned}
|\mathcal{W}_{t+1}-\mathcal{V}_{t+1}|&\leq C_{0,t}|\mathcal{W}_0-\mathcal{V}_0|+\Big|\sum_{n=0}^t C_{n+1,t} R'_{n+1}\Big|\\
&=
O(1/t)|\mathcal{W}_0-\mathcal{V}_0|+
%\sum_{n=0}^t O\Big(\frac{n}{t}\frac{1}{n^{2}}\Big)+
\sum_{n=0}^t O\Big(\frac{n}{t}\frac{1}{n^{1+\beta'}}\Big)\\
&=
O(1/t)|\mathcal{W}_0-\mathcal{V}_0|+
%\frac{1}{t}\sum_{n=0}^t O\Big(\frac{1}{n}\Big)+
\frac{1}{t}\sum_{n=0}^t O(n^{-\beta'})\,.
%\\
%&=O(1/t)|\mathcal{W}_0-\mathcal{V}_0|+
%%%O\Big(\frac{\ln(t)}{t}\Big)+
%\frac{1}{t}\sum_{n=0}^t O(n^{-\beta'})\,.
%%%\begin{cases}
%%%O(\ln(t)/t) \quad\text{if } \beta\geq 1\\
%%%O(1/t^\beta) \quad\text{if } \beta\in (0,1)\,.
%%%\end{cases}
\end{aligned}
$$

%%%%%%%%%%%%%%%%%%%%%%%%%%%%%%%%%%%%%%%%%%%%%%%%%%%%%%%%%%%%%
\section{The case of a reducible interaction matrix}\label{app-reducible-case}

When the interaction matrix $\Gamma$ is not irreducible, it is possible to decompose it in irreducible sub-matrices such that the union of the spectra of the sub-matrices coincides with the spectrum of the original matrix. In the following, we will describe an heuristic argument (also employed in \cite{iacopini-2020} and \cite{ale-cri-ghi-innovation-2023}), useful in order to detect the rate at which each $S_{t,h}$ grows along time in the case of a general matrix $\Gamma$.\\ 
 
\indent The dynamics that rules the vectorial process $\vS_t=(S_{t,1},\dots,S_{t,N})^\top$
 can be approximated (as $t\to +\infty$) by the linear system of (deterministic) differential equations 
\begin{equation*}
\dot{\mathbf{s}}(t) = \Gamma \frac{\mathbf{s}(t)}{t}
\end{equation*}
and hence we can say that $\vS_t\approx \vs(t)$ for $t\to +\infty$. 
By the change of variable $t=e^z$, we get
\begin{equation*}
\dot{\vs}(z) = \Gamma \vs(z)\,,
\end{equation*}
whose general solution is given by $\vs(z)=e^{\Gamma z}\mathbf{c}$. Now,
the term $e^{\Gamma z}$ can be expressed using the canonical Jordan form of the matrix $\Gamma$, 
so that we obtain
$$
\vs(z)=\sum_{k=1}^{r} e^{\gamma_k z} \sum_{i=0}^{p_k-1} z^i \mathbf{c}_i,
$$
where  $\gamma_1,\dots, \gamma_r$ are the distinct eigenvalues of $\Gamma$, $p_1,\dots,p_r$ are 
the sizes of the corresponding Jordan blocks and $\mathbf{c}_i$ are suitable vectors related to $\mathbf{c}$ and to 
the generalized eigenvectors of $\Gamma$. Indeed, 
we can write $\Gamma$ as $PJP^{-1}$, where $J$ is its canonical Jordan form and $P$ is a suitable invertible matrix of 
generalized eigenvectors.
Therefore, we have $e^{\Gamma z}=P e^{J z} P^{-1}$, where $e^{Jz}$ is a block matrix with blocks of the form
$e^{J_k z}$ with $J_k$ block in $J$. On the other hand, if $J_k =\gamma_k I + N_k$ is a generic Jordan 
block of $\Gamma$ with size $p_k$ and associated to the eigenvalue $\gamma_k$, we have
$$
e^{J_k z}= e^{\gamma_k z} e^{N_k z}=
e^{\gamma_k z} \sum_{i=0}^{p_k-1} \frac{z^i}{(i-1)!} N_k^i\, .
$$
Changing the variable from $z$ to $t$, we find 
\begin{equation}\label{eq-general-solution}
\vS_t\approx \vs(t) = 
\sum_{k=1}^{r} t^{\gamma_k} \sum_{i=0}^{p_k-1} \ln^i(t) \mathbf{c}_i
\end{equation}
and so the rate at which $S_{t,h}$ increases is given by the leading term in the expression
of $s_h(t)$. 

%%%%%%%%%%%%%%%%%%%%%%%%%%%%%%%%%%%%%%%%%%%%%%%%%%%%%%%%%%%%%%%%%%

\section{Choice of the threshold (robustness check)}\label{choice_threshold-app}
\begin{figure}[tbhp]
\centering
\includegraphics[height=8cm]{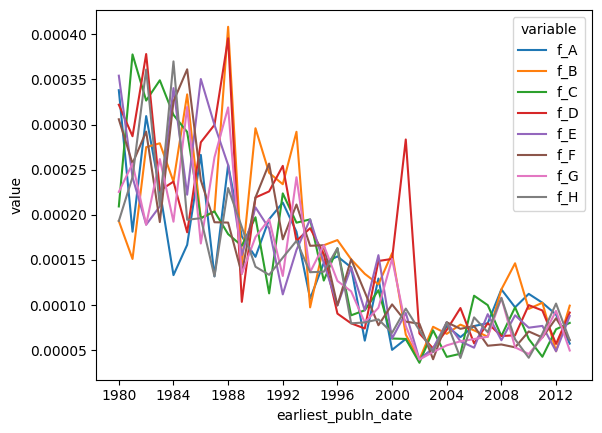}
\caption{Behavior of the mean index per category over the years.} 
\label{fig:mean-index}
\end{figure}
In this section we discuss the choice of the threshold $\tau$ that has been used in the data analysis of Section~\ref{sec:data_analysis} to define when a given patent $n$ can be considered a success for (or in) category $h$, that is, if and only if $I_{n,h}> \tau$.
Since we want to identify as  a ``success'' only patents that have an extraordinary impact on at least one category, the value of the threshold $\tau$ 
should be greater than the index value of the vast majority of patents in the data set.
However, this requirement is not so stringent in the sense that the percentage of patents with $(I_{n,h}>\tau)$ is already around $1\%$ for $\tau=0.1$ and is below $0.1\%$ for any  $\tau >0.5$ (see Figure~\ref{fig:mean-index} and Table~\ref{table:Table quantiles tau}  for further details). This means that the majority of patents in the data set have an index value very low (precisely, below $0.1$). 
This is in accordance with~\cite{squicciarini2013measuring}, who observed that only a very small 
subset of patents typically receives a
large number of forward citations and the mean value of the forward index decreases along time.\\
\begin{table}[tbhp]
  \centering
    \begin{tabular}{|c|ccccc|}
     \hline
category & $(I_{n,h}>0.1)\%$ & $(I_{n,h}>0.3)\%$ & $(I_{n,h}>0.5)\%$ & $(I_{n,h}>0.7)\%$& $(I_{n,h}>0.9)\%$\\ \hline
A    &      1.640\,\% &   0.185\,\% &  0.045\,\%  & 0.017\,\% &  0.008\,\%\\
B     &     2.440\,\% &   0.250\,\%  &  0.059\,\%  & 0.021\,\% &  0.008\,\%\\
C     &     1.180\,\% &   0.154\,\% &  0.042\,\%  & 0.017\,\% &  0.008\,\%\\
D     &     0.554\,\% &   0.1410\,\% &  0.031\,\%  & 0.017\,\% &  0.011\,\%\\
E     &     1.070\,\% &   0.149\,\% &  0.038\,\%  & 0.017\,\% &  0.009\,\%\\
F     &     1.420\,\% &   0.170\,\%   & 0.043\,\%  & 0.018\,\% &  0.008\,\%\\
G     &     1.450\,\% &   0.153\,\%  & 0.038\,\%  & 0.016\,\% &  0.008\,\%\\
H     &     1.350\,\% &   0.150\,\%   & 0.037\,\%  & 0.016\,\% &  0.008\,\%\\
 \hline
    \end{tabular}
\caption{For each category $h$, percentage of patents with a value of the index $I_{n,h}$ greater than $\tau=0.1,0.3,0.5,0.7,0.9$.}
 \label{table:Table quantiles tau}
\end{table}

\begin{figure}[tbhp]
\centering
\includegraphics[height=8cm]{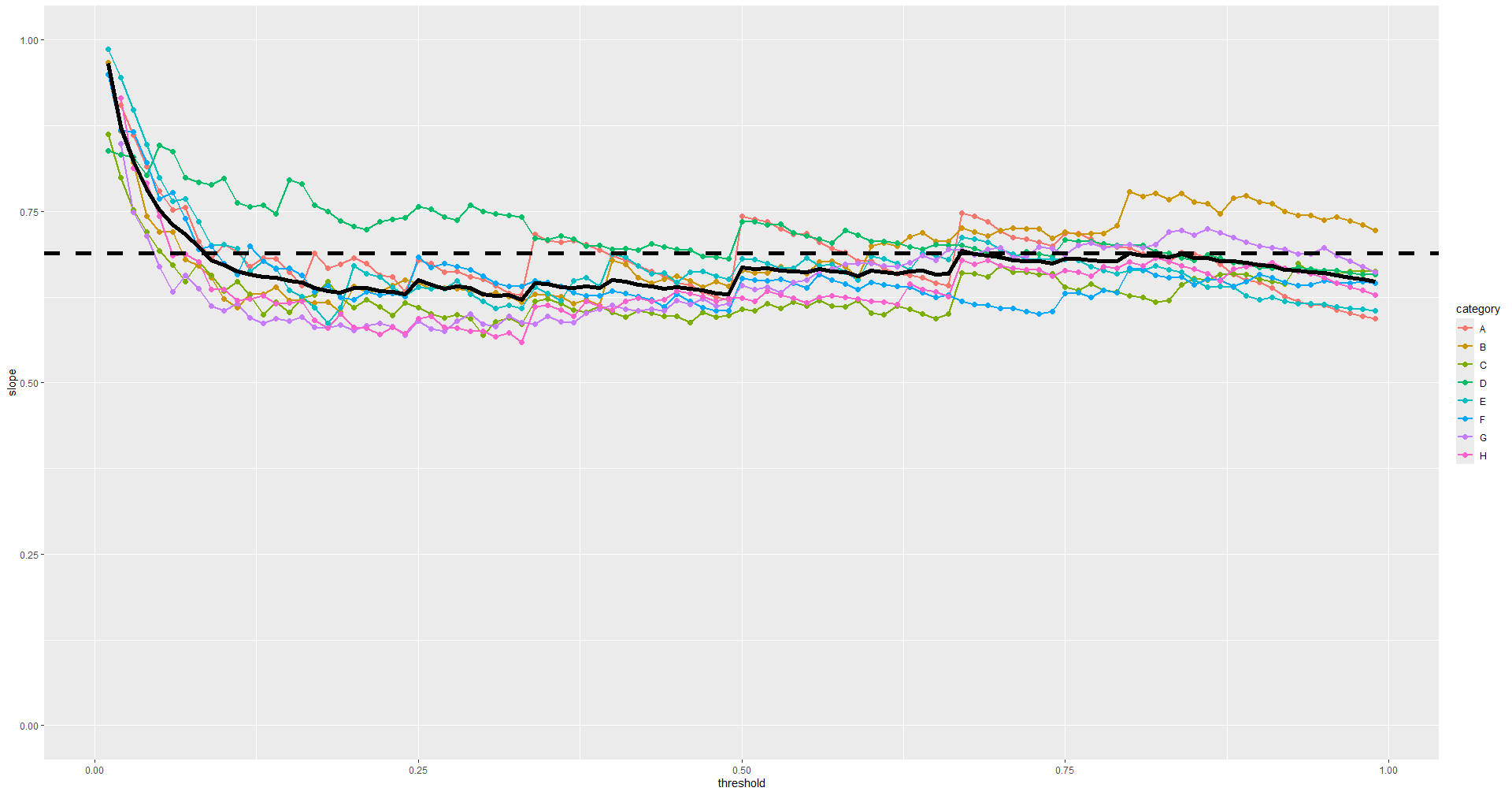}  
\caption{Different slopes estimated through linear regression on $S_{t,h}$, in $\log_{10}-\log_{10}$ scale, for each category $h$ (color line) and each value of the threshold $\tau$ (x-axis).
The solid black line indicates the common slope $\widehat{\gamma}^*$
estimated imposing the same slope for all categories.
The dashed horizontal black line indicates the common slope $\widehat{\gamma}^*=0.689$ estimated for $\tau=0.8$ in Section~\ref{sec:data_analysis}.}
\label{fig:slope_different_thresholds}  
\end{figure}
\begin{figure}[htp]
\centering
\includegraphics[height=8cm]{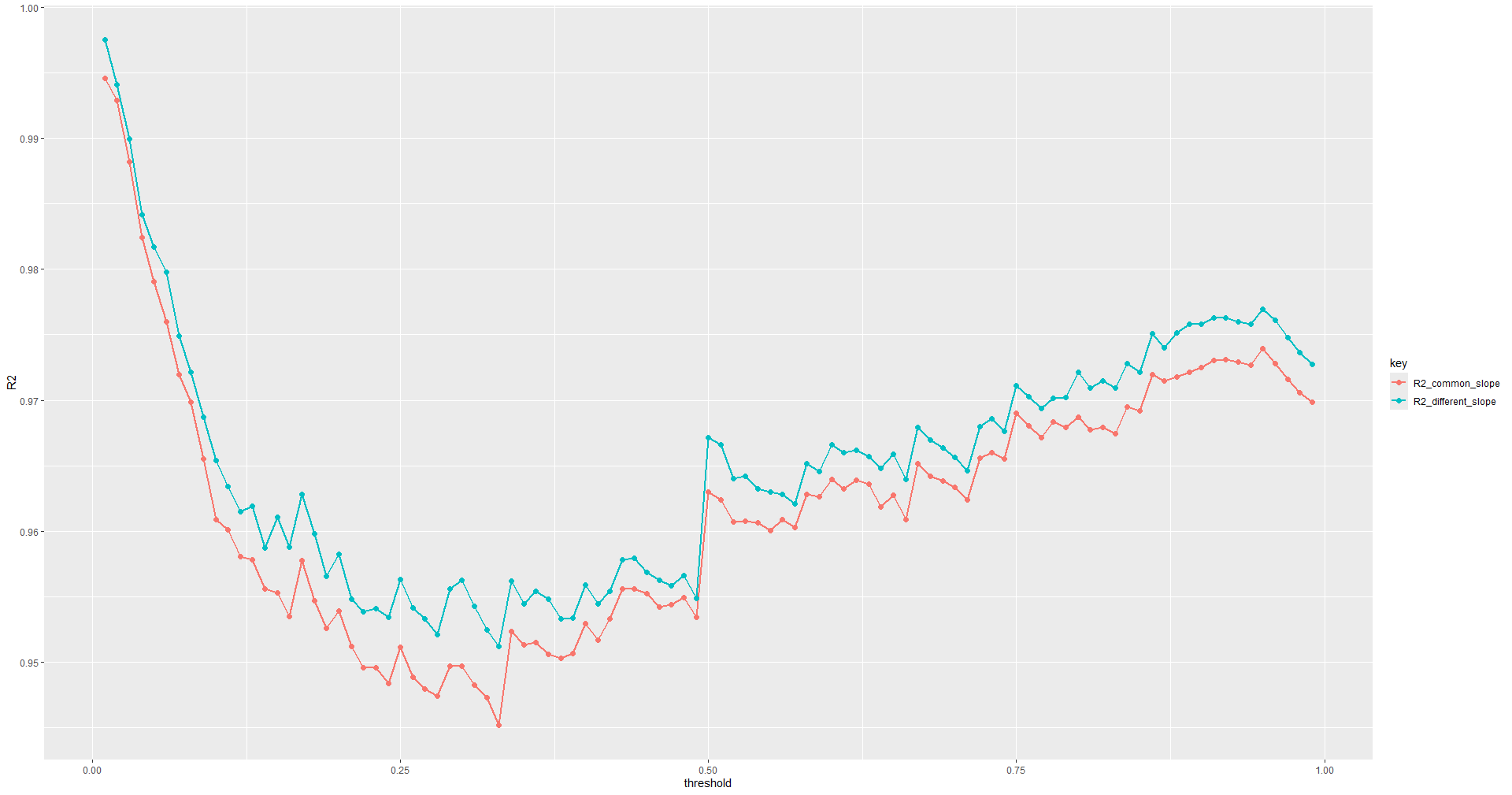}  
\caption{Goodness-of-fit index $R^2$ of the linear regression on $S_{t,h}$, in $\log_{10}-\log_{10}$ scale, for each value of the threshold $\tau$ (x-axis). The red line indicates the $R^2$ obtained imposing a common slope and
the blue line the $R^2$ obtained allowing the slopes to be different across categories. Both the lines are always very close to each other and always higher than $0.95$ for any value of $\tau$.}
\label{fig:R2_different_thresholds}  
\end{figure}

Then, to verify the robustness of the results of the paper with respect to the choice of the threshold $\tau$, we performed the data analysis presented in Section~\ref{sec:data_analysis} multiple times, each one using a different threshold value. In particular, we have performed a linear regression for every process $S_{t,h}$, in the $\log_{10}-\log_{10}$ scale, allowing the slopes to be different between categories and also imposing a common slope. The results are collected in
Figure~\ref{fig:slope_different_thresholds}, where we can see that
the variability presented by the slopes is essentially very similar for any value of $\tau>0.2$.
In addition, when we perform the linear regression that imposes the same slope for all categories, the estimated common slope $\widehat{\gamma}^*$ is always very close to the value $0.689$, that is the one estimated with $\tau=0.8$ in Section~\ref{sec:data_analysis}.
Finally, we have calculated the goodness-of-fit index $R^2$ obtained imposing a common slope and the one obtained allowing the slopes to be different across categories, and they are always very close to each other and always higher than 0.95 for any value of $\tau$ (see Figure~\ref{fig:R2_different_thresholds}).\\

\section{Econometric robustness check} \label{sec:rob_checks}

As a complementary robustness check, we estimate a reduced-form econometric specification that asks whether the empirical patterns emphasized by the model are also visible in an aggregated panel of technological categories over time. The purpose of the exercise is to verify, in a more standard econometric setting, whether success in a target category is positively associated with past success in the same category (path dependence) and with past success in the other categories (cross-fertilisation) -- both results suggested in the main analysis. 

To this end, we aggregate the patent-level success indicators \(X_{n,h}\) to the target-category--year level. Let \(Y_{h,\tau}\) denote the number of patents published in year \(\tau\) that qualify as successes in target category \(h\), and let \(N_{\tau}\) denote the total number of patents published in year \(\tau\). We then define the target-category success rate as
\[
r_{h,\tau}=\frac{Y_{h,\tau}}{N_{\tau}}.
\]
Using this rate-based measure, we estimate the following linear specification:
\[
r_{h,\tau}=\alpha_h+\lambda_{\tau}
+\beta_1 r_{h,\tau-1}
+\beta_2 \sum_{j\neq h} r_{j,\tau-1}
+\varepsilon_{h,\tau},
\]
where \(\alpha_h\) are category fixed effects and \(\lambda_{\tau}\) are year fixed effects. The coefficient \(\beta_1\) captures path dependence within category \(h\), while \(\beta_2\) captures cross-fertilisation from the other categories.

This formulation is motivated by the empirical structure of the paper. Since category-specific success is defined for each patent and each target category, the relevant aggregated outcome is the number of successes in target category \(h\) per patent opportunity in year \(\tau\). The rate specification therefore provides a natural reduced-form counterpart to the paper’s interpretation of success becoming rarer or more persistent relative to the scale of inventive opportunities. At the same time, it avoids the extreme sparsity of the patent-level binary specification.

Table~\ref{tab:robust_rate} reports the results. 

\begin{table}[H]
\centering
\caption{Econometric robustness check: target-category success rates}
\label{tab:robust_rate}
\begin{tabular}{lccc}
\hline\hline
 & (1) & (2) & (3) \\
 & Baseline & \makecell{Baseline +\\ FE} & \makecell{Baseline +\\ Charactersitcs}   \\
\hline
Lagged own-category success rate, \(r_{h,\tau-1}\) 
   & $0.162^{**}$  & $0.116^{**}$ & $0.127^{*}$  \\
   & (0.088) & (0.060) &  (0.067) \\

Lagged other-categories success rate, \(\sum_{j\neq h} r_{j,\tau-1}\)
    & $0.103^{***}$ & $0.113^{***}$ & $0.137^{***}$   \\
    & (0013) & (0.021) & (0.049)   \\

Category FE
    & No & \checkmark & \checkmark   \\

Year FE
    &No & \checkmark & \checkmark   \\
Characteristics
   &No & No & \checkmark   \\

Observations
    &272 & 272 & 252   \\

\(R^2\)
   & 0.620 & 0.771 & 0.759   \\
\hline\hline
\multicolumn{4}{p{0.88\linewidth}}{\footnotesize Notes: The dependent variable is the success rate in target category \(h\) and year \(\tau\), defined as \(r_{h,\tau}=Y_{h,\tau}/N_{\tau}\), where \(Y_{h,\tau}\) is the number of patents published in year \(\tau\) that qualify as successes in target category \(h\), and \(N_{\tau}\) is the total number of patents published in year \(\tau\). Robust standard errors in parentheses. \(^{*}p<0.10\), \(^{**}p<0.05\), \(^{***}p<0.01\).}
\end{tabular}
\end{table}

The coefficient on the lagged own-category success rate is positive and marginally significant in all specifications, while the coefficient on the lagged success rate in the other categories is positive and strongly significant. Hence, the reduced-form evidence is consistent with both path dependence and cross-fertilisation, with stronger support for the latter. We notice here that these findings should be interpreted as complementary to the main analysis of the paper: they do not replace the model-based results, but they show that the two central mechanisms highlighted by the theoretical framework are also detectable in a conventional econometric specification. \\
Column (1) regresses the target-category rate on the lag own and other categories' rates. Column (2) adds category and year fixed effects. Finally, Column (3) repeats the same linear rate specification as Column (2), but augments it with category-year characteristics. These controls summarize the observable composition of patenting activity within each technological category and year, including the average size of patent families, the average number of applicants, the average number of inventors, and the relative importance of different filing routes. The results leave the main picture broadly unchanged. The coefficient on the lagged own-category success rate remains positive, although it becomes less precisely estimated, while the coefficient on the lagged success rate in the other categories remains positive and statistically significant.

% \bibliographystyle{imsart-nameyear}
% \bibliography{esco} % if more than one, comma separated

\end{document}